\documentclass[12pt]{article}

\usepackage{amssymb}
\usepackage[dvips]{graphicx}
\usepackage{comment}

\makeatletter
 \@addtoreset{equation}{section}
 
\makeatother

\begin{document}


\begin{titlepage}

\renewcommand{\thefootnote}{\fnsymbol{footnote}}


\begin{flushright}
\begin{tabular}{l}
UTHEP-659\\
OIQP-13-12
\end{tabular}
\end{flushright}

\bigskip

\begin{center}
\Large \bf 
Multiloop Amplitudes of Light-cone Gauge Bosonic String Field Theory
in Noncritical Dimensions
\end{center}

\bigskip

\begin{center}
{\large Nobuyuki Ishibashi}${}^{a}$\footnote{e-mail:
        ishibash@het.ph.tsukuba.ac.jp}
and
{\large Koichi Murakami}${}^{b}$\footnote{e-mail:
        koichimurakami71@gmail.com}
\end{center}

\begin{center}
${}^{a}$\textit{Institute of Physics, University of Tsukuba,\\
Tsukuba, Ibaraki 305-8571, Japan}\\
\end{center}
\begin{center}
$^{b}$\textit{Okayama Institute for Quantum Physics,\\
      Kyoyama 1-9-1, Kita-ku, Okayama 700-0015, Japan} 
\end{center}

\bigskip

\bigskip

\bigskip

\begin{abstract}
We study the multiloop amplitudes of the light-cone gauge closed
bosonic string field theory for $d \neq 26$.
We show that the amplitudes can be recast into a BRST invariant form
by adding a nonstandard worldsheet theory 
for the longitudinal variables $X^{\pm}$ 
and the reparametrization ghost system.
The results obtained in this paper for bosonic strings
provide a first step towards the examination whether the dimensional 
regularization works for the multiloop amplitudes of the light-cone 
gauge superstring field theory.

\end{abstract}

\setcounter{footnote}{0}
\renewcommand{\thefootnote}{\arabic{footnote}}

\end{titlepage}

\section{Introduction}

Light-cone gauge string field theory~\cite{Kaku:1974xu,
Kaku:1974zz,Mandelstam:1973jk,Cremmer:1974ej}
is a formulation of string theory in which the unitarity of 
S-matrix is manifest.
In the light-cone gauge superstring field theory, regularization
is necessary to deal with the contact term
problem~\cite{Greensite:1986gv,Greensite:1987sm,Green:1987qu,
Greensite:1987hm,Wendt:1987zh}.
In Refs.~\cite{Baba:2009kr,Baba:2009ns,Baba:2009zm,Baba:2009fi,
Ishibashi:2010nq,Ishibashi:2011fy},
it has been proposed to employ dimensional regularization
to deal with the divergences in string field theory.
Being a completely gauge fixed formulation, 
the light-cone gauge NSR string field theory can be defined
in $d~(d\ne 10)$ dimensions.
By taking $d$ to be a large negative value,
the divergences of the amplitudes are regularized.\footnote{Precisely
    speaking, in order to deal with  the amplitudes
    involving the strings in the (R,NS) and the (NS,R) 
    sectors,
    we shift the Virasoro central charge of the system,
    rather than the spacetime dimensions 
    themselves~\cite{Ishibashi:2011fy}.
    This can be achieved
    by adding an extra conformal field
    theory with sufficiently large negative central 
    charge.}
Defining the amplitudes as analytic functions of $d$
and taking the $d \to 10$ limit, 
we obtain the amplitudes in critical dimensions.
So far, it has been verified that this dimensional regularization scheme
works for the closed string tree-level amplitudes and 
it has been shown that no divergences
occur in the limit $d\to 10$.
This implies that at least at tree level
we need not add contact interaction terms
to the string field theory action as counter terms.

It is obvious that what we should do next 
is to check
whether this regularization scheme
works for the multiloop amplitudes as well.
In this paper, as a first step towards this goal,
we study the multiloop amplitudes
of the light-cone gauge closed bosonic string field theory
in noncritical dimensions. 
We evaluate the multiloop amplitudes  and 
give them as integrals over moduli space of the Riemann surfaces 
corresponding to the Feynman diagrams for strings. 
We show that they can be rewritten
into a BRST invariant form using the conformal gauge
worldsheet theory.
This can be accomplished by adding
the longitudinal variables $X^{\pm}$
and the ordinary reparametrization $bc$ ghosts
in a  similar way to
the case of the tree-level amplitudes~\cite{Baba:2009ns}.
The worldsheet theory for
the longitudinal variables is
the conformal field theory
formulated in Ref.~\cite{Baba:2009ns}, which we refer to
as the $X^{\pm}$ CFT.

The organization of this paper is as follows.
In section~\ref{sec:lcamplitudes}, we consider the $h$-loop
$N$-string amplitudes for the light-cone gauge closed bosonic
string field theory in noncritical dimensions.
Such an amplitude corresponds to a light-cone string diagram
which is conformally equivalent to an $N$ punctured genus $h$ 
Riemann surface.
We present an expression of the amplitude as an integral over the moduli 
space of the Riemann surface. 
In section~\ref{sec:BRSTinv}, the $X^{\pm}$ CFT
on the higher genus Riemann surfaces is constructed.
We find that the prescription developed 
in the sphere case~\cite{Baba:2009ns} can be directly generalized to
the present case.
Introducing the reparametrization
ghost variables as well as the $X^{\pm}$ CFT, 
we rewrite amplitudes into those
of the BRST invariant formulation of strings
for $d \neq 26$ in the conformal gauge.
Section~\ref{sec:discussions} is devoted to summary and discussions.
In appendix~\ref{sec:thetaprime}, the definitions of 
the theta functions, the prime form and the Arakelov
Green's functions are presented.
In appendix~\ref{sec:Gamma}, the partition functions 
of the worldsheet theory on the string
diagrams are evaluated.
In appendix~\ref{sec:modular-prop}, we show that the amplitudes
of the light-cone gauge string theory are modular invariant
even in noncritical dimensions.
In appendix~\ref{sec:proof-ZLC-Zgh},
we present 
a derivation of an identity which is necessary
in section~\ref{sec:BRSTinv}
to rewrite the amplitudes into a BRST invariant form.

\section{Amplitudes of light-cone string field theory}
\label{sec:lcamplitudes}

\begin{figure}[h]
\begin{center}
\includegraphics[width=0.75\textwidth]{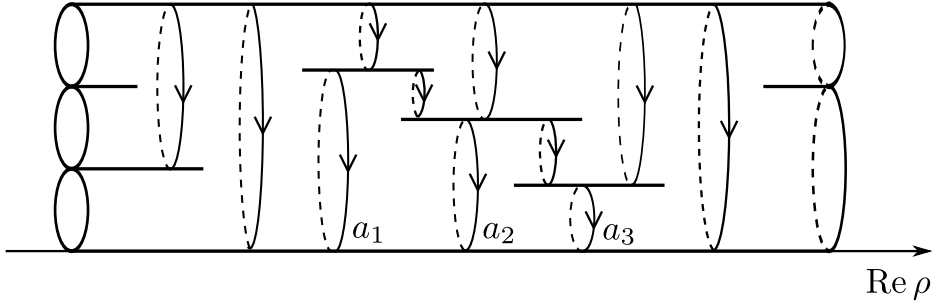}
\end{center}
\caption{A string diagram 
         with $3$ incoming, $2$ outgoing strings and $3$ loops.
        The cycles $\mathcal{C}_{\mathcal{I}}$
        going around the cylinders corresponding
        to the internal propagators are
        described. Among them,
        the cycle $a_{j}$ is the $a$ cycle for the $j$-th loop.}
\label{fig:stringdiagram}
\end{figure}

The light-cone gauge string field theory is defined
even for $d \neq 26$.
The action for the closed string field theory takes a simple form
consisting of a kinetic term and a cubic interaction term:
\begin{eqnarray}
S &=& \int dt \left[ \frac{1}{2} \int d1 d2
   \, \langle R(1,2) | \Phi (t) \rangle_{1}
   \left( i \frac{\partial}{\partial t}
          - \frac{L^{\mathrm{LC} (2)}_{0} + \tilde{L}^{\mathrm{LC} (2)}_{0}
                  - \frac{d-2}{12}}
                {\alpha_{2}}
  \right)
  | \Phi (t) \rangle_{2}
  \right.
\nonumber \\
&& \hphantom{\int dt [}
  \quad \left.  {}+ \frac{g}{6} \int
    d1d2d3 \, \langle V_{3} (1,2,3)|\Phi (t) \rangle_{1}
    |\Phi (t) \rangle_{2}
    |\Phi (t) \rangle_{3}
  \right].
\label{eq:sftaction}
\end{eqnarray}
Here $|\Phi \rangle_{r}$ is the string field,
$dr$ denotes the integration measure of the momentum
zero-modes given by
\begin{equation}
dr= \frac{\alpha_{r}d\alpha_{r}}{4\pi}
    \frac{d^{d-2}p_{r}}{(2\pi)^{d-2}}~,
\label{eq:dr}
\end{equation}
$\alpha_{r} = 2p^{+}_{r}$ is the string-length and
$L^{\mathrm{LC}(r)}_{0}$, $\tilde{L}^{\mathrm{LC}(r)}_{0}$
denote the zero-modes of
the transverse Virasoro generators for the $r$-th string.
The definitions of the reflector $\langle R(1,2)|$ and 
the three string vertex $\langle V_{3} (1,2,3)|$ are presented
in appendix~A of Ref.~\cite{Baba:2009ns}.
Starting from this action, we can evaluate the amplitudes perturbatively.
Each term in the expansion corresponds to a light-cone gauge 
Feynman diagram for strings. 
A typical $3$-loop $5$-string diagram is depicted in
Figure~\ref{fig:stringdiagram}.

\subsubsection*{Mandelstam mapping}

A Euclideanized  
$h$-loop $N$-string diagram is conformally equivalent
to an $N$ punctured genus $h$ Riemann surface $\Sigma$.
The light-cone diagram consists of cylinders. On each cylinder, 
one can introduce a complex coordinate $\rho$ whose real part 
coincides with the Euclideanized light-cone time $iX^+$ 
and imaginary part parametrizes the closed string at 
each time. 
The $\rho$'s on the cylinders are smoothly connected except at 
the interaction points and we get a complex coordinate $\rho$ 
on $\Sigma$. 
$\rho$ is not a good coordinate around the punctures and the 
interaction points on the light-cone diagram. 

$\rho$ can be expressed as an analytic function $\rho (z)$ 
in terms of a local complex coordinate $z$ on $\Sigma$.
As in the tree case, $\rho (z)$ is called the Mandelstam mapping. 
Let $Z_{r}$ $(r=1,\ldots,N)$ be
the $z$ coordinate of the puncture on $\Sigma$ which corresponds to
the $r$-th external leg of the light-cone diagram.
The Mandelstam mapping $\rho (z)$
can be determined by the two requirements:
\begin{enumerate}
\item The one-form
$d\rho =\partial \rho (z) dz$ 
should have simple poles at the punctures $Z_r$
with residues $\alpha_r$
and be non-singular everywhere else.\footnote{To be
precise, $d\rho$ is a meromorphic one-form on the Riemann surface
with $N$ marked points $Z_{r}$ $(r=1,\ldots,N)$,
and holomorphic one-form on the punctured Riemann surface,
which has these marked points removed.
\label{footnote:1}
}

\item $\mathop{\mathrm{Re}} \rho$ has to be globally defined on
$\Sigma$ because it is the light-cone time of the light-cone
string diagram.
$d\rho$ should
therefore have purely imaginary period around
any homology cycle.
\end{enumerate}
These two requirements uniquely fix 
the one-form $d\rho$~\cite{Giddings:1986rf,D'Hoker:1988ta} 
and the Mandelstam mapping $\rho (z)$ is obtained as
\begin{equation}
\rho (z) = \sum_{r=1}^{N} \alpha_{r}
  \left[ \ln E(z,Z_{r})
          - 2\pi i 
           \int^{z}_{P_{0}} \omega
           \frac{1}{\mathop{\mathrm{Im}} \Omega}
           \mathop{\mathrm{Im}} \int^{Z_{r}}_{P_{0}} \omega
  \right]~,
\qquad
\sum_{r=1}^{N} \alpha_{r} =0~,
\label{eq:Mandelstam}
\end{equation}
up to an additive constant independent of $z$.
Here 
$\omega =(\omega_{j})$ $(j=1,\ldots,h)$ denotes the canonical
basis of the holomorphic one-forms
and $\Omega=\left( \Omega_{jk} \right)$ is
the period matrix  on $\Sigma$, whose definitions are given
in appendix~\ref{sec:thetaprime}.
$P_{0}$ is an arbitrary point on $\Sigma$, which we 
take as the base point of the Abel-Jacobi map,
and  
$E(z,w)$ is the prime form~\cite{Fay1973,Mumford1984}
defined in eq.(\ref{eq:primeform}).
Since the one-form $d\rho$ has $N$ simple poles,
$d\rho$  has $2h-2+N$ simple zeros, 
which we denote by $z_{I}$
$(I=1,\ldots,2h-2+N)$.
They correspond to the interaction points of 
the light-cone string diagram.

On the light-cone string diagram, the flat
metric
\begin{equation}
ds^{2} =d\rho d\bar{\rho}
\label{eq:metric-rho}
\end{equation}
is chosen as usual, 
which is referred to as the Mandelstam metric.
In terms of a local coordinate $z$ on $\Sigma$,
it takes the form
\begin{equation}
ds^{2} = \left|\partial \rho (z)\right|^{2} dz d\bar{z}~.
\label{eq:metric-z}
\end{equation}
This metric is singular at the punctures $z=Z_r$
and the interaction points 
$z=z_{I}$.
For later use,
as is done for the tree-level light-cone string diagrams,
we introduce the local coordinate $w_{r}$ around the puncture
at $z=Z_{r}$ defined as
\begin{equation}
w_{r} (z) 
 = \exp \left[ \frac{1}{\alpha_{r}} 
                \left( \rho (z) -\rho(z_{I^{(r)}}) \right)
        \right]~,
\label{eq:localcoord}
\end{equation}
where $z_{I^{(r)}}$ denotes the interaction point on the $z$-plane
where the $r$-th external string interacts.

\subsubsection*{Amplitudes}

It is straightforward to calculate the amplitudes by the 
old-fashioned perturbation theory 
starting from the action (\ref{eq:sftaction}) and 
Euclideanize the time integrals. 
An $h$-loop $N$-string amplitude
is given as an integral over the moduli space
of the string diagram~\cite{D'Hoker:1987pr} as
\begin{equation}
\mathcal{A}_{N}^{(h)}
 = (ig)^{2h-2+N}
   C
   \int [dT] [\alpha d\theta ] [d\alpha ]
    \, 
F_{N}^{(h)}~,
\label{eq:amplitude}
\end{equation}
where $\int [dT] [\alpha d\theta ] [d\alpha ]$ denotes the integration over
the moduli parameters and $C$ is the combinatorial factor. 
In each channel, the integration measure is given as
\begin{equation}
\int [dT] [\alpha d\theta ] [d\alpha ]
=
\prod_{a=1}^{2h-3+N} \left( -i \int_0^\infty dT_{a} \right)
\prod_{A=1}^{h}\int\frac{d\alpha_A}{4\pi}
\prod_{\mathcal{I}=1}^{3h-3+N}
\left( |\alpha_\mathcal{I}| 
       \int^{2\pi}_{0} \frac{d\theta_{\mathcal{I}}}{2\pi}
\right).
\end{equation}
Here $T_a$'s are heights of the cylinders corresponding to  
internal lines,\footnote{Heights of the cylinders in a light-cone 
  diagram are constrained 
  so that only $2h-3+N$ of them can be varied independently.}
$\alpha_A$'s denote the string-lengths corresponding to 
the $+$ components of the loop momenta 
and $\alpha_{\mathcal{I}}$'s and $\theta_{\mathcal{I}}$'s are 
the string-lengths and the twist angles
for the internal propagators.

The integrand $F_{N}^{(h)}$ can be described as a correlation
function in the light-cone gauge worldsheet theory
for the transverse variables $X^{i}$ $(i=1,\ldots,d-2)$.
It takes the form \cite{Baba:2009ns}
\begin{equation}
F_{N}^{(h)} 
= (2\pi)^{2} \delta \left( \sum_{r=1}^{N} p^{+}_{r} \right)
             \delta \left(\sum_{r=1}^{N} p^{-}_{r} \right)
  \mathop{\mathrm{sgn}}
    \left( \prod_{r=1}^{N} \alpha_{r}
    \right)
 \left( Z^{\mathrm{LC}} \right)^{\frac{d-2}{24}}
 \left\langle \prod_{r=1}^{N} V_{r}^{\mathrm{LC}}
 \right\rangle^{X^{i}}~.
\label{eq:FN1}
\end{equation}
Here 
$V^{\mathrm{LC}}_{r} \equiv V^{\mathrm{LC}}_{r} (w_{r}=0,\bar{w}_{r}=0)$
denotes the light-cone vertex operator for the $r$-th external
leg located at the origin of the local coordinate $w_{r}$
in eq.(\ref{eq:localcoord}), 
and $\left\langle \prod_{r=1}^{N} V_{r}^{\mathrm{LC}}
\right\rangle^{X^{i}}$ is the expectation value of these operators 
in the worldsheet theory for the transverse variables $X^i$. 
$(Z^\mathrm{LC})^{\frac{d-2}{24}}$ 
denotes the partition function of the worldsheet theory 
on the light-cone diagram. 
The factor $\mathop{\mathrm{sgn}}\left( \prod_{r=1}^{N} \alpha_{r}\right)$ 
comes from the peculiar form of the measure of $\alpha_r$ 
in eq.(\ref{eq:dr}) and 
our convention for the phase of the vertex $\langle V_3|$. 
The partition function 
$Z^{\mathrm{LC}}$ for $d=26$
is calculated in appendix~\ref{sec:Gamma} 
to be
\begin{equation}
Z^{\mathrm{LC}}
 = \frac{1}{(32\pi^{2})^{4h}} e^{-2(h-1)c} e^{2\delta (\Sigma)}
   \prod_{r} \left[ e^{-2\mathop{\mathrm{Re}} \bar{N}^{rr}_{00}}
                    (2g^{\mathrm{A}}_{Z_{r}\bar{Z}_{r}})^{-1}
                    \alpha_{r}^{-2} \right]
   \prod_{I} \left[ \left| \partial^{2} \rho (z_{I}) \right|^{-1}
                    2g^{\mathrm{A}}_{z_{I}\bar{z}_{I}} \right].
\label{eq:ZLCexpression}
\end{equation}
Here $g^{\mathrm{A}}_{z\bar{z}}$ denotes the Arakelov metric
defined in appendix~\ref{sec:thetaprime}. The quantities $c$, 
$\delta (\Sigma )$ and  
$\bar{N}^{rr}_{00}$ are defined in 
eqs.(\ref{eq:chi}), (\ref{eq:Faltings}) and (\ref{eq:Neumann00-holo}) 
respectively.

When the $r$-th external state is of the form
\begin{equation}
\alpha^{i_{1} (r)}_{-n_{1}} \cdots
\tilde{\alpha}^{\tilde{\imath}_{1} (r)}_{-\tilde{n}_{1}} \cdots
 \left|p^{-},p^{i}\right\rangle_{r}~,
\label{eq:LCstate}
\end{equation}
the vertex operator $V^{\mathrm{LC}}_{r}$ becomes
\begin{eqnarray}
V^{\mathrm{LC}}_{r} 
&=& \alpha_{r}
  \oint_{0} \frac{dw_{r}}{2\pi i}
         i\partial X^{i_{1}} (w_{r}) w_{r}^{-n_{1}}
  \cdots
  \oint_{0} \frac{d\bar{w}_{r}}{2\pi i} 
         i\bar{\partial} X^{\tilde{\imath}_{1}}
         (\bar{w}_{r})
         \bar{w}_{r}^{-\tilde{n}_{1}} \cdots
\nonumber \\
&& \ \times
  e^{-p_{r}^{-}\tau_{0}^{(r)}}
  e^{i p^{i}_{r} X^{i}} \left(w_{r}=0,\bar{w}_{r}=0\right)~,
\label{eq:VLC}
\end{eqnarray}
in which the operators are normal ordered,
$w_{r}$ denotes the local coordinate introduced 
in eq.(\ref{eq:localcoord}) and
$\tau^{(r)}_{0} = \mathop{\mathrm{Re}} \rho \left(z_{I^{(r)}}\right)$.
The on-shell and the level-matching conditions require that
\begin{equation}
\frac{1}{2} \left( -2p^{+}_{r}p^{-}_{r} + p^{i}_{r}p^{i}_{r} \right)
 + \mathcal{N}_{r} = \frac{d-2}{24}~,
\qquad
\mathcal{N}_{r} \equiv \sum_{j} n_{j} = \sum_{j} \tilde{n}_{j}~.
\label{eq:onshell-cond}
\end{equation}
We note that the definition of
$V_{r}^{\mathrm{LC}}$ 
is independent of
the choice of the worldsheet metric to define the theory and 
so is the expectation value 
$\left\langle \prod_{r=1}^{N} V_{r}^{\mathrm{LC}}
  \right\rangle^{X^{i}}$.
The expectation value 
$\left\langle \prod_{r=1}^{N} V_{r}^{\mathrm{LC}}
 \right\rangle^{X^{i}}$ can be 
expressed in terms of the Green's functions on $\Sigma$. 
Therefore we can express the integrand~(\ref{eq:FN1}) 
in terms of various quantities on  $\Sigma$
defined in appendix \ref{sec:thetaprime}. 

\subsubsection*{Comments}
Before closing this section, several comments are in order. 

\begin{enumerate}
\item
Using the Mandelstam mapping, 
$Z^{\mathrm{LC}}$ can be considered as the partition function
on the surface endowed with the Mandelstam metric~(\ref{eq:metric-z})
and  be written as
\begin{equation}
Z^{\mathrm{LC}} 
 = \left(
     Z^{X} \left[ \frac{1}{2} |\partial \rho (z) |^{2} 
           \right]
   \right)^{24}~,
\label{eq:ZLC-def}
\end{equation}
where 
\begin{equation}
Z^{X}[g_{z\bar{z}}] 
\equiv \left( 
  \frac{8\pi^{2} \det' \bigtriangleup_{g_{z\bar{z}}}}
       {\int dz \wedge d\bar{z} \sqrt{g}}
        \right)^{-\frac{1}{2}}~,
\label{eq:def-ZX}
\end{equation}
and $\bigtriangleup_{g_{z\bar{z}}}
 = - 2 g^{z\bar{z}} \partial_{z} \partial_{\bar{z}}$
denotes the scalar Laplacian for conformal gauge metric 
$ds^2=2g_{z\bar{z}}dzd\bar{z}$.
{}From the expression (\ref{eq:ZLC-def}), we get
\begin{equation}
Z^{\mathrm{LC}}
  = Z^{X} [g_{z\bar{z}}^\mathrm{A}]^{24}
      e^{-\Gamma 
            \left[g_{z\bar{z}}^\mathrm{A},\,\ln |\partial \rho|^{2}
            \right] }~,
\end{equation}
where
$\Gamma \left[ g_{z\bar{z}}^\mathrm{A},\, \ln |\partial \rho|^{2}
          \right]$
is the Liouville action 
\begin{equation}
\Gamma \left[ g_{z\bar{z}}^\mathrm{A},\,\ln |\partial \rho|^{2} \right]
 = -\frac{24}{48\pi} \int dz \wedge d\bar{z} \, i
     \left( \partial \chi \bar{\partial} \chi
             + g_{z\bar{z}}^\mathrm{A} R^\mathrm{A} \chi \right)~,
\label{eq:Gamma-Liouville}
\end{equation}
with $\chi (z,\bar{z}) = \ln |\partial \rho (z)|^{2}
                   - \ln (2g_{z\bar{z}}^\mathrm{A})$.

In Refs.~\cite{Sonoda:1987ra,D'Hoker:1989ae}, 
$\Gamma \left[ g_{z\bar{z}}^\mathrm{A},\, \ln |\partial \rho|^{2}
          \right]$
is calculated by directly evaluating the Liouville action 
(\ref{eq:Gamma-Liouville}). 
In order to do so,
one needs to regularize the divergences coming from the singularities 
of the Mandelstam metric~(\ref{eq:metric-z}).
In Ref.~\cite{Sonoda:1987ra} 
a Weyl invariant but reparametrization noninvariant regularization 
is employed and 
$e^{- \Gamma \left[ g_{z\bar{z}}^\mathrm{A},\, 
                    \ln |\partial \rho|^{2}
             \right]}$
is evaluated to be
\begin{equation}
e^{-2(h-1)c}
   \prod_{r} \left[
                    (2g_{Z_{r}\bar{Z}_{r}}^\mathrm{A})^{-1}
                    \alpha_{r}^{2} \right]
   \prod_{I} \left[ \left| \partial^{2} \rho (z_{I}) \right|^{-2}
                    2g_{z_{I}\bar{z}_{I}}^\mathrm{A} \right]~.
\label{eq:Sonoda}
\end{equation}
Our result for $Z^{\mathrm{LC}}$ implies that 
$e^{- \Gamma \left[ g_{z\bar{z}}^\mathrm{A},\, 
                    \ln |\partial \rho|^{2}
             \right]}$ 
is the one given in eq.(\ref{eq:expressionI-Gamma}),
which includes extra factors of 
$e^{-2\mathop{\mathrm{Re}} \bar{N}^{rr}_{00}}$ and
$\left| \partial^{2} \rho (z_{I}) \right|$ 
compared with eq.(\ref{eq:Sonoda}). 
These factors make 
$\Gamma \left[ g_{z\bar{z}}^\mathrm{A},\, \ln |\partial \rho|^{2}
          \right]$ 
both Weyl and reparametrization invariant 
as it should be.\footnote{It will be possible to get such factors 
  by using the more intricate 
  regularization method in Ref.~\cite{Mandelstam:1985ww}. }
The regularization employed in Ref.~\cite{Sonoda:1987ra} does not cause 
any problems in deriving the equivalence of the light-cone amplitudes
and the covariant ones in the critical dimensions, 
but it is not appropriate for the noncritical case. 

\item
In the light-cone gauge expression (\ref{eq:amplitude}) of the amplitudes, 
the moduli parameters $T, \alpha , \theta$ are modular invariant,
and the integration region covers the moduli space 
only once~\cite{Giddings:1986rf}. 
Being a function of these parameters, the 
integrand $F_N^{(h)}$ should also be invariant under the 
modular transformations. 
However, the explicit form (\ref{eq:ZLCexpression}) 
of $Z^\mathrm{LC}$ and 
the correlation function 
$\left\langle \prod_{r=1}^{N} V_{r}^{\mathrm{LC}}
 \right\rangle^{X^{i}}$ 
are given in terms of the quantities such as the 
theta functions which depend on the choice of the cycles $a_j, b_j$. 
As a consistency check, it is possible to show that these quantities 
are invariant under the modular transformations and do not depend on 
the choice of these cycles. The details are given in 
appendix~\ref{sec:modular-prop}.
\end{enumerate}

\section{BRST invariant form of the amplitudes}
\label{sec:BRSTinv}

In this section, we would like to rewrite the integrand $F^{(h)}_{N}$
of the amplitudes,
given in eq.(\ref{eq:FN1}), into the correlation
function of the worldsheet theory for strings 
in the conformal gauge.
By doing so, we will show that there exists a BRST invariant 
formulation in the conformal gauge corresponding to 
the string field theory in the noncritical dimensions. 
All these have been done for the tree-level amplitudes 
in Ref.~\cite{Baba:2009ns}.

\subsection{$X^{\pm}$ CFT}
\label{sec:XpmCFT}

In order to get the BRST invariant formulation, we need to introduce the 
longitudinal variables $X^{\pm}$ and the reparametrization $bc$ ghosts. 
The worldsheet theory of $X^{\pm}$ for $d\neq 26$ is constructed 
and called 
$X^\pm$ CFT~\cite{Baba:2009ns}.
In this subsection, we would like to consider the $X^\pm$ CFT 
on the Riemann surface $\Sigma$ of genus $h$.

For $d=26$, the longitudinal variables $X^{\pm}$ are introduced in the form 
of the following path integral:
\begin{equation}
\int \left[dX^+dX^-\right]_{\hat{g}_{z\bar{z}}}
e^{-S^\pm_{d=26}}
\prod_{r=1}^{N} e^{-ip^{+}_{r} X^{-}} (Z_{r},\bar{Z}_{r})
\prod_{s=1}^{M} e^{-ip^{-}_{s} X^{+}} (z_{s},\bar{z}_{s})
\,,
\label{eq:pathpm}
\end{equation}
where
\begin{equation}
S^\pm_{d=26}
=
-\frac{1}{4\pi}\int dz\wedge d\bar{z}\,i 
  \left( \partial X^+\bar{\partial}X^-
         + \partial X^{-} \bar{\partial} X^{+}
  \right) \,.
\end{equation}
Here we take a worldsheet metric $ds^2=\hat{g}_{z\bar{z}}dzd\bar{z}$ to define 
the path integral measure. 
For $d=26$, we do not have to worry about the choice of $\hat{g}_{z\bar{z}}$. 

Since the action for $X^\pm$ is not bounded below, we need to take 
the integration contour of $X^\pm$ carefully to define the integral. 
We decompose the variable $X^\pm$ as
\begin{equation}
X^\pm (z,\bar{z})
=
X_\mathrm{cl}^\pm (z,\bar{z})+x^{\pm}+\delta X^\pm (z,\bar{z})~,
\end{equation}
where 
$X_\mathrm{cl}^\pm (z,\bar{z})$ are solutions 
to the equations of motion with the source terms\footnote{The
 delta function $\delta^{2} (z)$
 in eq.(\ref{eq:Xpmcl}) is normalized so that
 $\int dz \wedge d\bar{z} \,\delta^{2} (z) =1$.}
\begin{eqnarray}
\partial\bar{\partial}X^{+}_{\mathrm{cl}} (z,\bar{z})
&=&
-i\sum_r p_r^+(-2\pi i)\delta^2(z-Z_r)\,,
\nonumber
\\
\partial\bar{\partial}X^{-}_{\mathrm{cl}} (z,\bar{z})
&=&
-i\sum_s p_s^-(-2\pi i)\delta^2(z-z_s)\,, 
\label{eq:Xpmcl}
\end{eqnarray}
and $x^{\pm} + \delta X^{\pm} (z,\bar{z})$ are the fluctuations
around the solutions
with
\begin{equation}
\int dz\wedge d\bar{z} \sqrt{\hat{g}} \delta X^{\pm} =0~.
\end{equation}
The integrals over $X^\pm$ are expressed as those over 
$x^\pm$ and $\delta X^\pm$. 
In eq.(\ref{eq:pathpm}), we take the integration contours of $x^\pm$ and 
$\delta X^+-\delta X^-$ to be 
along the real axis and that of 
$\delta X^++\delta X^-$ to be along the imaginary axis. 
Then eq.(\ref{eq:pathpm}) becomes well-defined and is evaluated to be
\begin{equation}
(2\pi )^2 \delta
\left( \sum_r p_r^+ \right) \delta \left( \sum_s p_s^- \right)
\prod_{s=1}^{M} e^{-ip^{-}_{s} X_\mathrm{cl}^{+}} (z_{s},\bar{z}_{s})
Z^{X}[\hat{g}_{z\bar{z}}]^2~. 
\label{eq:vevFX+L}
\end{equation}
Here we can see that the insertion 
$\prod_{s=1}^{N} e^{-ip^{-}_{s} X^{+}} (z_{s},\bar{z}_{s})$ 
in the path integral (\ref{eq:pathpm}) is replaced by its classical value 
$\prod_{s=1}^{N} e^{-ip^{-}_{s} X^{+}_\mathrm{cl}} (z_{s}.\bar{z}_{s})$. 
From eq.(\ref{eq:Xpmcl}), we can take
\begin{equation}
X_\mathrm{cl}^+ (z,\bar{z})
=
-\frac{i}{2} 
 \left(\rho \left(z  \right)
       +\bar{\rho} \left( \bar{z} \right)
 \right)
\; ,
\end{equation}
the right hand side of which coincides with the the Lorentzian time of 
the light-cone string diagram. 
As will be explained in the next subsection, 
one can relate the DDF vertex operators and the light-cone gauge 
ones using this fact. 
Therefore, by considering the path integral (\ref{eq:pathpm}), 
one can introduce the longitudinal variables 
essentially satisfying the light-cone gauge 
conditions. 
Multiplying the  path integral for the transverse variables $X^{i}$
by the path integral~(\ref{eq:pathpm}) for $X^{\pm}$
and that for the $bc$ ghosts, 
we are able to get the path integral for the 
conformal gauge worldsheet theory.

For $d\neq 26$, we should specify the metric on the worldsheet to 
define the path integral. 
In the light-cone gauge formulation, the natural choice is
\begin{equation}
ds^2=-4\partial X^+\bar{\partial}X^+dzd\bar{z}\,.
\end{equation}
This choice is meaningful only when $\partial X^+\bar{\partial}X^+$ 
possesses a 
reasonable expectation value. In our case, the expectation value 
coincides with the Mandelstam metric~(\ref{eq:metric-z}). 
The path integral measure defined with such a metric 
can be given as
\begin{equation}
\left[ dX^{i} dX^{\pm} db d\tilde{b} dc d\tilde{c}
\right]_{-4\partial X^+\bar{\partial}X^+}
 = \left[ dX^{i} dX^{\pm} db d\tilde{b} dc d\tilde{c}
   \right]_{\hat{g}_{z\bar{z}}} 
   e^{-\frac{d-26}{24} 
       \Gamma  \left[\hat{g}_{z\bar{z}},\, 
       \ln \left( -4\partial X^+\bar{\partial}X^+\right)
              \right]}~.
\end{equation}
Thus an extra dependence on the variable $X^+$ comes from the 
path integral measure.
Including it into the action for $X^\pm$ variables, 
the action $S^\pm$ becomes
\begin{equation}
S^\pm \left[ \hat{g}_{z\bar{z}} \right]
=
-\frac{1}{4\pi}\int dz\wedge d\bar{z}\,i 
  \left( \partial X^+\bar{\partial}X^- + \partial X^{-} \bar{\partial}X^{+}
  \right)
+
\frac{d-26}{24} 
       \Gamma  \left[\hat{g}_{z\bar{z}},\, 
       \ln \left( -4\partial X^+\bar{\partial}X^+\right)
              \right]\,.
\end{equation}
In order to relate the light-cone gauge formulation 
and the conformal gauge one,
we should consider
\begin{equation}
\int \left[dX^+dX^-\right]_{\hat{g}_{z\bar{z}}}
e^{-S^\pm \left[ \hat{g}_{z\bar{z}} \right]}
\prod_{r=1}^{N} e^{-ip^{+}_{r} X^{-}} (Z_{r},\bar{Z}_{r})
\prod_{s=1}^{M} e^{-ip^{-}_{s} X^{+}} (z_{s},\bar{z}_{s})~.
\label{eq:dne26}
\end{equation}
This path integral can be evaluated easily as follows. 
Using the prescription for the contour of the integration, 
it is straightforward to prove 
\begin{eqnarray}
0
&=&
\int \left[dX^+dX^-\right]_{\hat{g}_{z\bar{z}}}
e^{-S^\pm_{d=26}}
\nonumber
\\
& &
\quad \times
\prod_{r=1}^{N} e^{-ip^{+}_{r} X^{-}} (Z_{r}.\bar{Z}_{r})
\prod_{s=1}^{M} e^{-ip^{-}_{s} X^{+}} (z_{s}.\bar{z}_{s})
\prod_{p}\partial \delta X^+ (z_p) 
\prod_{q}\bar{\partial} \delta X^+ (\bar{z}_{q})~.
\label{eq:partial0}
\end{eqnarray}
It follows that substituting 
$X^+ = X^+_\mathrm{cl} + x^{+} + \delta X^+$ 
into the term 
$\Gamma  \left[\hat{g}_{z\bar{z}},\, 
     \ln \left( -4\partial X^+\bar{\partial}X^+\right)
              \right]$ 
of the action $S^\pm$ 
in eq.(\ref{eq:dne26}) and expanding it in terms of $\delta X^+$, 
we get 
\begin{eqnarray}
\lefteqn{
\left\langle 
\prod_{r=1}^{N} e^{-ip^{+}_{r} X^{-}} (Z_{r}.\bar{Z}_{r})
\prod_{s=1}^{M} e^{-ip^{-}_{s} X^{+}} (z_{s}.\bar{z}_{s})
\right\rangle^{X^\pm}_{\hat{g}_{z\bar{z}}}
}
\nonumber
\\
& &
\equiv
Z^{X}[\hat{g}_{z\bar{z}}]^{-2}
\int \left[dX^+dX^-\right]_{\hat{g}_{z\bar{z}}}
e^{-S^\pm \left[ \hat{g}_{z\bar{z}} \right]}
\prod_{r=1}^{N} e^{-ip^{+}_{r} X^{-}} (Z_{r},\bar{Z}_{r})
\prod_{s=1}^{M} e^{-ip^{-}_{s} X^{+}} (z_{s},\bar{z}_{s})
\nonumber
\\
& &
=
(2\pi )^2\delta\left(\sum_sp_s^-\right)\delta\left(\sum_rp_r^+\right)
\prod_se^{-p_s^-\frac{\rho +\bar{\rho}}{2}}(z_s,\bar{z}_s)
   \, e^{-\frac{d-26}{24}
       \Gamma  \left[\hat{g}_{z\bar{z}},\, \ln |\partial \rho|^{2}
              \right]}~.
\label{eq:vevFX+}
\end{eqnarray}
Taking 
$\Gamma  \left[\hat{g}_{z\bar{z}},\, \ln |\partial \rho|^{2}\right]$ 
to be the one given in eq.(\ref{eq:ZLC-critical}), 
it is possible to calculate various correlation functions 
of $X^\pm$ from eq.(\ref{eq:vevFX+}). 
One can show that the energy-momentum tensor of the $X^\pm$ CFT 
satisfies the Virasoro algebra with the central charge 
$28-d$~\cite{Baba:2009ns}. 
Therefore the worldsheet theory
for $X^\pm ,X^i, b, c, \tilde{b}, \tilde{c}$ 
is a CFT with vanishing central charge.


\subsection{BRST invariant form of amplitudes}

Using eq.(\ref{eq:vevFX+}), we find that 
the product of the light-cone vertex operators $V_r^\mathrm{LC}$
each of which corresponds to the light-cone state~(\ref{eq:LCstate})
is expressed as the expectation
value of that of the DDF vertex operators in the $X^{\pm}$ CFT:
\begin{eqnarray}
&&
(2\pi)^{2} \delta \left(\sum_{r=1}^{N} p^{+}_{r} \right)
           \delta \left( \sum_{r=1}^{N} p^{-}_{r} \right)
 \prod_{r=1}^{N} V^{\mathrm{LC}}_{r}
\nonumber \\
&&= \prod_{r=1}^{N} 
     \left( \alpha_{r} 
            e^{2 \mathop{\mathrm{Re}} \bar{N}^{rr}_{00}}
     \right)
e^{\frac{d-26}{24}
       \Gamma  \left[\hat{g}_{z\bar{z}},\, \ln |\partial \rho|^{2}
              \right]}
 \left\langle \prod_{r=1}^{N} \left[
     V^{\mathrm{DDF}}_{r} (Z_{r},\bar{Z}_{r})
     e^{\frac{d-26}{24} \frac{i}{p^{+}_{r}} X^{+}}
       \left(z_{I^{(r)}},\bar{z}_{I^{(r)}} \right)
     \right]
   \right\rangle^{X^{\pm}}_{\hat{g}_{z\bar{z}}}.
\nonumber \\
\label{eq:VLC-DDF}
\end{eqnarray}
Here $V^{\mathrm{DDF}}_{r}$ is the DDF vertex operator given by
\begin{equation}
V^{\mathrm{DDF}}_{r} (z,\bar{z})
 = A^{i_{1} (r)}_{-n_{1}} (z) \cdots
   \tilde{A}^{\tilde{\imath}_{1} (r)}_{-\tilde{n}_{1}} (\bar{z})
   \cdots
    e^{-ip^{+}_{r}X^{-} 
        -i\left(p^{-}_{r}-\frac{\mathcal{N}_{r}}{p^{+}_{r}}
                +\frac{d-26}{24} \frac{1}{p^{+}_{r}}
          \right) X^{+}
        + ip^{i}_{r} X^{i}}
     (z,\bar{z})~,
\label{eq:VDDF}
\end{equation}
with the DDF operator $A^{i (r)}_{-n}$  for the $r$-th string
defined as
\begin{equation}
A^{i(r)}_{-n} (z)
 = \oint_{z} \frac{dz'}{2\pi i} \, i\partial X^{i} (z')
   e^{-i \frac{n}{p^{+}_{r}} X^{+}_{L} (z')}~,
\label{eq:DDFop}
\end{equation}
and $\tilde{A}^{\tilde{\imath} (r)}_{-\tilde{n}}$
similarly given for the anti-holomorphic sector. 
The operators in eq.(\ref{eq:VDDF}) are normal ordered,
and $X^{+}_{L}(z)$ in eq.(\ref{eq:DDFop})
denotes the holomorphic part
of $X^{+}(z,\bar{z})$.

Substituting eq.(\ref{eq:VLC-DDF}) into eq.(\ref{eq:FN1})
with the worldsheet metric $\hat{g}_{z\bar{z}}$ 
taken to be $g_{z\bar{z}}^\mathrm{A}$,
we find that the integrand $F^{(h)}_{N}$ 
of the amplitude~(\ref{eq:amplitude})
is expressed as
\begin{eqnarray}
F^{(h)}_{N} 
&\propto&  
    \prod_{r=1}^{N}
   \left( \alpha_{r} e^{2 \mathop{\mathrm{Re}} \bar{N}^{rr}_{00}} \right)
    e^{-\Gamma  \left[g_{z\bar{z}}^\mathrm{A},\, \ln |\partial \rho|^{2}
              \right]}
   Z^{X} [g_{z\bar{z}}^\mathrm{A}]^{-2}
   \nonumber
   \\
& &
\quad
\times
    Z^{X} [g_{z\bar{z}}^\mathrm{A}]^{d}
   \left\langle
     \prod_{r=1}^{N}
                 \left[ V^{\mathrm{DDF}}_{r} (Z_{r},\bar{Z}_{r})
                        e^{\frac{d-26}{24} \frac{i}{p^{+}_{r}} X^{+}}
                           \left(z_{I^{(r)}},\bar{z}_{I^{(r)}} \right)
                 \right]
   \right\rangle^{X^{\mu}}_{g^{\mathrm{A}}_{z\bar{z}}},
\label{eq:FN2}
\end{eqnarray}
where $\langle \cdots \rangle^{X^{\mu}}_{\hat{g}_{z\bar{z}}}$ 
denotes
the correlation function in the combined system 
of the worldsheet theory for $X^{i}$
$(i=1,\ldots,d-2)$ and the $X^{\pm}$ CFT
with the metric $\hat{g}_{z\bar{z}}$.

We will further rewrite eq.(\ref{eq:FN2}) 
by introducing the ghost variables. 
It is possible to show the following identity:
\begin{eqnarray}
&&
    \prod_{r=1}^{N}
   \left( \alpha_{r} e^{2 \mathop{\mathrm{Re}} \bar{N}^{rr}_{00}} \right)
    e^{-\Gamma  \left[g_{z\bar{z}}^\mathrm{A},\, \ln |\partial \rho|^{2}
              \right]}
   Z^{X} [g_{z\bar{z}}^\mathrm{A}]^{-2}
\nonumber \\
&& = \mbox{const.}
\int\left[dbd\tilde{b}dcd\tilde{c}\right]_{g_{z\bar{z}}^\mathrm{A}}
e^{-S^{bc}}
\prod_{r=1}^{N}c\tilde{c}(Z_r, \bar{Z}_r)
\prod_{K=1}^{6h-6+2N}
   \left[ \int dz \wedge d\bar{z} \,i \left( \mu_{K}b
          +   
           \bar{\mu}_{K}\tilde{b}\right)
    \right].~~~~~~~~
\label{eq:ghostZX}
\end{eqnarray}
Here 
$S^{bc}$ is the action for the $bc$ ghosts,
$\mu_K$ $(K=1,\cdots ,6h-6+2N)$ denote the Beltrami differentials 
for the moduli parameters $T, \alpha , \theta$, and
$\mbox{const.}$ indicates a constant
independent of the moduli parameters.
Eq.(\ref{eq:ghostZX})
is derived in appendix~\ref{sec:proof-ZLC-Zgh}.

Substituting eq.(\ref{eq:ghostZX}) into eq.(\ref{eq:FN2}), 
we eventually get
\begin{eqnarray}
\mathcal{A}_{N}^{(h)}
 &\sim&
\int [dT] [ d\alpha] [\alpha d\theta] 
\nonumber
\\
& &
\times
\int \left[dX^\mu dbd\tilde{b}dcd\tilde{c}
     \right]_{g_{z\bar{z}}^\mathrm{A}}
e^{-S^{X^{i}} -S^{\pm} \left[ g^{\mathrm{A}}_{z\bar{z}} \right]
   -S^{bc}}
\prod_{K=1}^{6h-6+2N}
   \left[ \int dz\wedge d\bar{z} \, i \left(\mu_{K}b
          + 
            \bar{\mu}_{K}\tilde{b}\right)
   \right]
\nonumber
\\
& &
\hphantom{ 
          \int \left[dX^\mu dbd\tilde{b}dcd\tilde{c}
     \right]_{g_{z\bar{z}}^\mathrm{A}}
    }
\times
\prod_{r=1}^{N}
\left[
c\tilde{c}V^{\mathrm{DDF}}_{r} (Z_{r},\bar{Z}_{r})
e^{\frac{d-26}{24} \frac{i}{p^{+}_{r}} X^{+}}
\left(z_{I^{(r)}},\bar{z}_{I^{(r)}} \right)
\right],
\label{eq:BRSTinvariant}
\end{eqnarray}
where $S^{X^{i}}$ denotes the action for the worldsheet
theory of the transverse variables $X^{i}$.
This form of the amplitude is BRST invariant. 
The vertex operators $c\tilde{c}V^{\mathrm{DDF}}_{r}$ are BRST invariant 
on the mass shell, the insertions 
$e^{\frac{d-26}{24} \frac{i}{p^{+}_{r}} X^{+}}$
at interaction points are BRST invariant~\cite{Baba:2009ns}
and the BRST variation of the antighost insertion yields the 
total derivative with respect to the moduli parameter. 
Eq.(\ref{eq:BRSTinvariant}) is a generalization of 
the results in Refs. \cite{D'Hoker:1987pr, Sonoda:1987ra} 
to the noncritical case.

\section{Summary and discussions}
\label{sec:discussions}

In this paper, we have studied the multiloop amplitudes
of the light-cone gauge
bosonic string field theory in noncritical dimensions.
The amplitudes are expressed as integrals over the moduli space 
of the Riemann surfaces corresponding to the light-cone diagrams. 
We have constructed the worldsheet theory for the $X^{\pm}$ variables 
on the higher genus surfaces.
It has been shown that the multiloop amplitudes
of the light-cone string theory in noncritical dimensions
can be rewritten into BRST invariant ones 
of the conformal gauge worldsheet theory
consisting of the theory for the transverse
variables $X^{i}$,
the $X^{\pm}$ CFT and the reparametrization
$bc$ ghost system.

We may be able to construct the gauge invariant 
string field theory based on the conformal gauge worldsheet theory
mentioned above.
Since we have constructed the CFT on the worldsheet of
the light-cone string diagram,
this string field theory should possess 
the joining-splitting type interaction.
Such a theory is expected to be 
a version formulated in noncritical dimensions 
of the $\alpha = p^{+}$ HIKKO
theory~\cite{Kugo:1992md,Hata:1986jd,Hata:1986kj}.

Since the on-shell condition is given by 
eq.(\ref{eq:onshell-cond}), 
we can regularize the infrared behavior of the amplitudes by taking 
$d$ to be negative and large. 
The divergences of multiloop amplitudes of string theory are 
infrared divergences and we expect that 
our formulation works as a regularization of them. 
The expression for the multiloop amplitudes in the light-cone gauge
formalism will not be so useful for practical calculations. 
We need to know the coordinates $z_I$ of the interaction points 
to evaluate the amplitudes, which is technically impossible for almost 
all the cases. 
Moreover, the parametrization of the moduli space is not holomorphic. 
What we intend to do is not finding a way of calculations, but showing that 
the amplitudes can be deduced from a simple string field theory
action with only cubic interaction term. 

Bosonic string theory itself is not so interesting anyway, because 
of the existence of tachyon. 
In order to discuss tachyon free theory, 
we need to supersymmetrize the analyses in this paper and
investigate
whether the dimensional regularization scheme proposed in 
Refs.~\cite{Baba:2009kr,Baba:2009ns,Baba:2009zm,Baba:2009fi,
Ishibashi:2010nq,Ishibashi:2011fy} 
works for the multiloop amplitudes in
the light-cone gauge NSR superstring field theory.
Using the supersheet 
technique~\cite{Berkovits:1985ji,Berkovits:1987gp,
Berkovits:1988xq,Aoki:1990yn},
it will be possible to relate the results of the light-cone gauge formalism to 
those of the covariant formalism using the super Riemann surfaces
\cite{Witten:2012bh, Witten:2013cia, Witten:2012bg, Witten:2012ga}.

\section*{Acknowledgements}

We are grateful to F.~Sugino for discussions.
N.I.\ would like to acknowledge the hospitality of 
Okayama Institute for Quantum Physics
and K.M.\ would like to thank the hospitality of
Particle Theory Group at University of Tsukuba,
where part of this work was done.
This work was supported in part by 
Grant-in-Aid for Scientific Research~(C) (20540247),
(23540332) and 
(25400242) from MEXT.


\appendix

\section{Theta functions, prime form and Arakelov Green's function}
\label{sec:thetaprime}

In this appendix, we explain various quantities defined on a
genus $h$ Riemann surface $\Sigma$, 
which are necessary to express the multiloop amplitudes 
of the light-cone gauge 
string field theory.\footnote{The mathematical background 
relevant for string perturbation theory is
reviewed in Ref.~\cite{D'Hoker:1988ta}}

In the usual way, we choose on $\Sigma$ a canonical basis
$\{a_{j}, b_{j}\}$ $(j=1,\ldots, h)$ of homology cycles.
Let $\omega = (\omega_{j})=(\omega_{1},\ldots,\omega_{h})$
be the dual basis of holomorphic one-forms on $\Sigma$:
\begin{equation}
\oint_{a_{j}} \omega_{k} = \delta_{jk}~,
\quad
\oint_{b_{j}} \omega_{k} = \Omega_{jk}~,
\label{eq:norm-omega}
\end{equation}
where $\Omega=(\Omega_{jk})$ is the period matrix,
which is a symmetric $h\times h$ complex matrix
with positive definite imaginary part,
$\mathop{\mathrm{Im}} \Omega >0$.

\subsubsection*{Theta functions}

With the period matrix $\Omega$ in eq.(\ref{eq:norm-omega}),
any point $\delta \in \mathbb{C}^{h}$ can be uniquely 
expressed in terms of two $\mathbb{R}^{h}$-vectors as
\begin{equation}
\delta = \delta^{\prime} \Omega + \delta^{\prime\prime}~,
\qquad
\delta^{\prime},\delta^{\prime\prime} \in \mathbb{R}^{h}~.
\label{eq:characteristics}
\end{equation}
The notation
$[\delta] 
  = {\delta^{\prime} \atopwithdelims[] \delta^{\prime\prime}}$
is used to represent the point $\delta \in \mathbb{C}^{h}$
in eq.(\ref{eq:characteristics}).
The theta function with characteristics
$[\delta]
  ={\delta^{\prime} \atopwithdelims[] \delta^{\prime\prime}}$
is defined by
\begin{eqnarray}
\theta[\delta] (\zeta|\Omega)
 &=&
    \sum_{n\in \mathbb{Z}^{h}} 
       e^{2\pi i \left[ \frac{1}{2} (n+\delta^{\prime})
                                    \Omega (n+\delta^{\prime})
                        + (n+\delta^{\prime}) 
                          (\zeta+\delta^{\prime\prime})
                 \right]}
\nonumber \\
  &=& e^{2\pi i \left[ \frac{1}{2}\delta^{\prime}
                            \Omega \delta^{\prime}
                       + \delta^{\prime}
                         (\zeta+\delta^{\prime\prime})
                \right]}
      \, \theta(\zeta+\delta^{\prime\prime} 
                   +\delta^{\prime} \Omega|\Omega)~,
\label{eq:theta-char}
\end{eqnarray}
where 
$\theta(\zeta|\Omega) = \theta [0](\zeta|\Omega)$.
$\theta [\delta]  (\zeta|\Omega)$
is a quasi-periodic function on the Jacobian variety
$J(\Sigma) 
  = \mathbb{C}^{h}/(\mathbb{Z}^{h}+\mathbb{Z}^{h}\Omega)$
of the Riemann surface $\Sigma$
and transforms as
\begin{equation}
\theta [\delta]
  (\zeta+ m+n\Omega|\Omega)
= e^{2\pi i m \delta^{\prime}}
  e^{-2\pi i n \delta^{\prime\prime}}
  e^{-\pi i n\Omega n - 2\pi i n \zeta}
  \, \theta [\delta] (\zeta|\Omega)
\label{eq:periodicity-char1}
\end{equation}
for $m,n \in \mathbb{Z}^{h}$.
We note that from eq.(\ref{eq:theta-char}) we have
\begin{equation}
\left|\theta[\zeta](0|\Omega)\right|
 = e^{-\pi \mathop{\mathrm{Im}} \zeta
            \frac{1}{\mathop{\mathrm{Im}} \Omega}
            \mathop{\mathrm{Im}} \zeta}
   \left| \theta (\zeta |\Omega) \right|~.
\label{eq:abs-theta-char}
\end{equation}

It is immediate from the definition (\ref{eq:theta-char}) that
\begin{equation}
\theta {\delta^{\prime}+m 
        \atopwithdelims[] \delta^{\prime\prime} + n}
      (\zeta|\Omega)
 = e^{2\pi i \delta^{\prime} n} 
      \theta {\delta^{\prime} \atopwithdelims[] 
               \delta^{\prime\prime}} (\zeta|\Omega)
\label{eq:shift-characteristics}
\end{equation}
for $m,n \in \mathbb{Z}^{h}$.
Thus $\theta [\delta]$ only changes its phase 
if $\delta^{\prime}$ and $\delta^{\prime\prime}$
are shifted by integral vectors.
The case in which 
$\delta^{\prime},\delta^{\prime\prime} \in
  \left(\mathbb{Z}/(2\mathbb{Z}) \right)^{h}$ 
is important.
In this situation,
$[\delta] 
  = {\delta^{\prime} \atopwithdelims[] \delta^{\prime\prime}}$
is referred to as the spin structure, and we have
\begin{equation}
\theta [\delta] (-\zeta|\Omega)
 = (-1)^{4 \delta^{\prime} \delta^{\prime\prime}} 
      \theta [\delta] (\zeta|\Omega)~.
\label{eq:spinstructure}
\end{equation}
It follows that
$\theta [\delta](\zeta|\Omega)$ is an even or odd function
depending on whether $4\delta^{\prime}\delta^{\prime\prime}$ is
an even or odd integer. 
$[\delta]$ is accordingly referred to as
the even spin structure or the odd spin structure.

\subsubsection*{Prime form} 

Let 
$[s]={s^{\prime} \atopwithdelims[] s^{\prime\prime}}$ 
be an odd spin structure.
The prime form $E(z,w)$ is defined~\cite{Fay1973,Mumford1984} as
\begin{equation}
E(z,w) =
   \frac{ \theta[s]
          \left(\left. \int^{z}_{w} \omega \right|\Omega\right)}
        {h_{s} (z) h_{s} (w)}~,
\label{eq:primeform}
\end{equation}
where 
\begin{equation}
h_{s} (z) = \sqrt{\sum_{j=1}^{h} 
  \frac{\partial \theta [s]}{\partial \zeta_{j}} (0|\Omega)
  \omega_{j} (z)}
\end{equation}
is a section of the spin bundle corresponding to $[s]$.
The prime form $E(z,w)$ can be regarded as
a $\left(-\frac{1}{2},0\right)$ form in each variable
on the universal covering of $\Sigma$,
whose transformation laws can be obtained
from eq.(\ref{eq:periodicity-char1}) as follows:
When $z$ is moved around $a_{j}$ cycle once,
$E(z,w)$ is invariant up to a sign;
whereas when $z$ is moved around $b_{j}$ cycle once, it transforms as
\begin{equation}
E(z,w) \mapsto
   \pm
   e^{-\pi i \Omega_{jj} - 2\pi i \int^{z}_{w} \omega_{j}}
   E(z,w)~.
\label{eq:E-around-b}
\end{equation}
$E(z,w)$ satisfies $E(z,w) = - E(w,z)$, 
and for $z \sim w$ it behaves as
\begin{equation}
E(z,w) = (z-w) + \mathcal{O}\left( (z-w)^{3} \right)~.
\label{eq:Eshort}
\end{equation}

\subsubsection*{Arakelov metric and Arakelov Green's function}

Let us define $\mu_{z\bar{z}}$ as
\begin{equation}
\mu_{z\bar{z}} 
\equiv
   \frac{1}{2h}
   \omega(z) 
   \frac{1}{\mathop{\mathrm{Im}} \Omega}
   \bar{\omega} (\bar{z})~.
\label{eq:Bergman}
\end{equation}
We note that
\begin{equation}
\int_{\Sigma} dz \wedge d\bar{z} \, i \mu_{z\bar{z}} =1~,
\label{eq:norm-Bergman}
\end{equation}
which follows from
\begin{equation}
\int_{\Sigma} \omega_{j} \wedge \bar{\omega}_{k}
 = -2i \mathop{\mathrm{Im}} \Omega_{jk}~.
\label{eq:omega-volume}
\end{equation}
The Arakelov metric on $\Sigma$,
\begin{equation}
ds_{\mathrm{A}}^{\; 2} = 2g^{\mathrm{A}}_{z\bar{z}} dz d\bar{z}~,
\label{eq:Arakelov}
\end{equation}
is defined so that its scalar curvature
$R^{\mathrm{A}} 
  \equiv -2g^{\mathrm{A}z\bar{z}} \partial \bar{\partial}
          \ln g^{\mathrm{A}}_{z\bar{z}}$ satisfies
\begin{equation}
g^{\mathrm{A}}_{z\bar{z}} R^{\mathrm{A}} 
= -8\pi (h-1) \mu_{z\bar{z}}~.
\label{eq:ArakelovR}
\end{equation}
This condition determines $g^{\mathrm{A}}_{z\bar{z}}$
only up to an overall constant,
which we will choose later.

The Arakelov Green's function
$G^{\mathrm{A}} (z,\bar{z};w,\bar{w})$ with respect to
the Arakelov metric is defined to satisfy
\begin{eqnarray}
&&-\partial_{z} \partial_{\bar{z}}
    G^{\mathrm{A}}(z,\bar{z};w,\bar{w})
   =2\pi \delta^{2} (z-w) -2\pi \mu_{z\bar{z}}~,
\nonumber \\
&& \int_{\Sigma} dz \wedge d\bar{z} \, i \mu_{z\bar{z}} 
   G^{\mathrm{A}} (z,\bar{z};w,\bar{w})
=0~.
\label{eq:G-B}
\end{eqnarray}
One can obtain a more explicit form of
$G^{\mathrm{A}}(z,\bar{z};w,\bar{w})$
by solving eq.(\ref{eq:G-B}) for $G^{\mathrm{A}}(z,\bar{z};w,\bar{w})$.
Using eq.(\ref{eq:Eshort}), we have
\begin{equation}
\partial_{z}\partial_{\bar{z}} \ln F(z,\bar{z};w,\bar{w})
 = -2\pi i \delta^{2} (z-w) - 2\pi h \mu_{z\bar{z}}~,
\label{eq:diffeq-propagator}
\end{equation}
where $F(z,\bar{z};w,\bar{w})$ is the
$\left(-\frac{1}{2},-\frac{1}{2}\right)
   \times \left( -\frac{1}{2},-\frac{1}{2}\right)$ form
on $\Sigma \times \Sigma$ defined as
\begin{equation}
F(z,\bar{z};w,\bar{w})
 = \exp \left[ -2\pi 
                \mathop{\mathrm{Im}} \int^{z}_{w} \omega
                \frac{1}{\mathop{\mathrm{Im}}\Omega}
                \mathop{\mathrm{Im}} \int^{z}_{w} \omega
        \right]
  \left| E(z,w) \right|^{2}~.
\label{eq:propagator2}
\end{equation}
Putting eqs.(\ref{eq:diffeq-propagator}) and (\ref{eq:ArakelovR})
together,
we find that $G^{\mathrm{A}}(z,\bar{z};w,\bar{w})$ is given by
\begin{equation}
G^{\mathrm{A}}(z,\bar{z};w,\bar{w})
 = -\ln F(z,\bar{z};w,\bar{w})
    -\frac{1}{2} \ln \left( 2 g^{\mathrm{A}}_{z\bar{z}}\right)
    -\frac{1}{2} \ln \left( 2 g^{\mathrm{A}}_{w\bar{w}} \right)~,
\label{eq:G-F-g-g}
\end{equation}
up to an additive constant independent of $z,\bar{z}$ and $w,\bar{w}$.
This possible additive constant can be absorbed into the ambiguity
in the overall constant of 
$g^{\mathrm{A}}_{z\bar{z}}$ mentioned above.
It is required that eq.(\ref{eq:G-F-g-g}) holds exactly
as it is~\cite{Dugan:1987qe,Sonoda:1987ra,D'Hoker:1989ae}.
This implies that
\begin{equation}
2g^{\mathrm{A}}_{z\bar{z}} 
= \lim_{w\to z}
  \exp \left[ -G^{\mathrm{A}}(z,\bar{z};w,\bar{w}) - \ln |z-w|^{2} \right]~,
\label{eq:gA-expGA}
\end{equation}
and the overall constant of $g^{\mathrm{A}}_{z\bar{z}}$
is, in principle, determined by the second relation
in eq.(\ref{eq:G-B}).


\subsubsection*{Mandelstam mapping}
Here we illustrate several properties of 
the Mandelstam mapping~(\ref{eq:Mandelstam}). 

The divisor
$D_{d\rho} = \sum_{I=1}^{2h-2+N} z_{I} - \sum_{r=1}^{N} Z_{r}$
of the one-form $d\rho$
satisfies
\begin{equation}
\sum_{I=1}^{2h-2+N} \int^{z_{I}}_{P_{0}} \omega
 - \sum_{r=1}^{N} \int^{Z_{r}}_{P_{0}} \omega
= 2 \Delta \qquad \pmod{\mathbb{Z}^{h}+\mathbb{Z}^{h}\Omega}~.
\label{eq:divisor-rho}
\end{equation}
Here $\Delta$ is the vector of
Riemann constants for $P_{0}$, which is defined
in 
$J(\Sigma)$.
Its $j$-th component $\Delta_{j}$ is given by
\begin{equation}
\Delta_{j} 
= -\frac{\Omega_{jj}}{2} +\frac{1}{2}
  + \sum_{k\neq j}  \oint_{a_{k}} \omega_{k} (P') 
                  \int_{P_{0}}^{P'} \omega_{j}~.
\label{eq:Riemannconst}
\end{equation}

{}From the singular behavior of the Mandelstam metric~(\ref{eq:metric-z}), 
one can find that
it satisfies the differential equation
\begin{equation}
-\partial \bar{\partial} 
  \ln \left| \partial \rho (z) \right|^{2}
 = i 2\pi \left( \sum_{I} \delta^{2} (z-z_{I})
                  - \sum_{r} \delta^{2} (z-Z_{r})
          \right)~.
\end{equation}
This can be solved as
\begin{equation}
\left| \partial \rho (z) \right|^{2}
  = 2 g_{z\bar{z}}^\mathrm{A} e^{\chi (z,\bar{z})}~,
\label{eq:metric-z-Arakelov}
\end{equation}
with
\begin{equation}
\chi (z,\bar{z}) 
  = \sum_{r=1}^{N} G^\mathrm{A}(z;Z_{r}) 
     - \sum_{I=1}^{2h-2+N} G^\mathrm{A}(z;z_{I})
    +c~,
\label{eq:chi}
\end{equation}
where $c$ is a constant independent of $z,\bar{z}$
but may depend on moduli.
We here and henceforth suppress the anti-holomorphic coordinate
dependence of $G^{\mathrm{A}}$ for brevity of notation.
Looking at the behaviors of eq.(\ref{eq:metric-z-Arakelov})
around $z \sim z_{I}$ and $z \sim Z_{r}$,
one finds that
\begin{eqnarray}
\left|\partial^{2} \rho (z_{I}) \right|^{2}
 &=& \left( 2 g^{\mathrm{A}}_{z_{I}\bar{z}_{I}} \right)^{2}
   \exp \left[ -\sum_{J\neq I} G^\mathrm{A} (z_{I} ; z_{J})
            + \sum_{r} G^\mathrm{A} (z_{I} ; Z_{r} )
            +c \right]~,
\nonumber \\
\left| \alpha_{r} \right|^{2}
 &=& \exp \left[ -\sum_{I} G^\mathrm{A} (z_{I} ; Z_{r})
   +\sum_{s \neq r} G^\mathrm{A} (Z_{r} ; Z_{s}) +c
    \right]~,
\label{eq:alpha-Arakelov}
\end{eqnarray}
and thus
\begin{eqnarray}
\prod_{I} \left|\partial^{2} \rho (z_{I}) \right|^{2}
&=& e^{(2h-2+N)c}
    \prod_{I} \left( 2 g_{z_{I}\bar{z}_{I}}^\mathrm{A}
              \right)^{2}
    \exp \left[-2\sum_{I<J}G^\mathrm{A} (z_{I} ; z_{J})
             + \sum_{I,r} G^\mathrm{A}(z_{I} ; Z_{r} )
   \right],
\nonumber \\
\prod_{r} \left| \alpha_{r} \right|^{2}
 &=& e^{Nc} 
     \exp \left[ -\sum_{I,r} G^\mathrm{A} (z_{I} ; Z_{r})
                 +2\sum_{r<s} G^\mathrm{A} (Z_{r} ;Z_{s})
     \right].~~~~
\label{eq:partial2rhozI}
\end{eqnarray}

\section{Evaluation of $Z^{\mathrm{LC}}$}
\label{sec:Gamma}

In this appendix, we will evaluate the partition
function $(Z^{\mathrm{LC}})^{\frac{d-2}{24}}$ for the transverse coordinates
$X^{i}$. 
In the following, we consider the case $d=26$ to get the 
partition function $Z^{\mathrm{LC}}$. 

\begin{figure}[htbp]
\begin{center}
\includegraphics[width=0.5\textwidth,keepaspectratio=true]{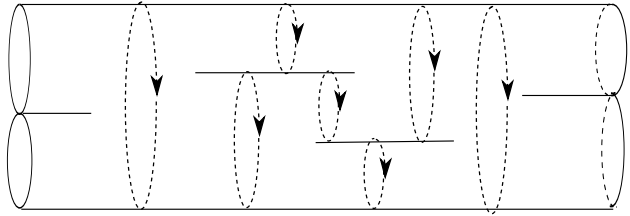}
\end{center}
\caption{The contours $C_{\mathcal{I}}$.}
\label{fig:CmathcalI}
\end{figure}

$Z^{\mathrm{LC}}$ can be obtained by integrating 
the change $\delta \ln Z^{\mathrm{LC}}$
under the variation of moduli parameters,
as in the tree case~\cite{Baba:2009ns}. 
Since $Z^{\mathrm{LC}}$ is the partition function, 
if we vary the lengths and the twist angles of 
the internal propagators
of the light-cone diagram, 
the change of $Z^{\mathrm{LC}}$
is given in terms of the expectation values of the 
Hamiltonian and the rotation generator as
\begin{equation}
\delta\ln Z^{\mathrm{LC}} 
 =  \sum_{\mathcal{I}}\delta \mathcal{T}_{\mathcal{I}}
      \oint_{C_{\mathcal{I}}}\frac{d\rho}{2\pi i}
      \left\langle T_{\rho\rho}^{\mathrm{tr}}\right\rangle^{X^{i}}
 +\mathrm{c.c.}\,.
\label{eq:deltaZLC1}
\end{equation}
Here $\mathcal{I}$ labels the internal lines of 
the light-cone diagram $\Sigma$ and 
$C_{\mathcal{I}}$
denotes the contour going around it as depicted 
in Figure~\ref{fig:CmathcalI}. 
$\mathcal{T}_{\mathcal{I}}$ is defined as
\begin{equation}
\mathcal{T}_{\mathcal{I}}
=
T_{\mathcal{I}}+i\alpha_{\mathcal{I}}\theta_{\mathcal{I}}\,,
\end{equation}
where
$T_{\mathcal{I}}$ denotes the length of the $\mathcal{I}$-th
internal line and $\alpha_{\mathcal{I}}$, $\theta_{\mathcal{I}}$ 
denote the string-length and the twist angle 
for the propagator. 
$\mathop{\mathrm{Re}}\delta \mathcal{T}_{\mathcal{I}}$'s 
should satisfy some linear constraints 
so that the variation corresponds to that 
of the shape of a light-cone diagram. 
$\left\langle T_{\rho\rho}^{\mathrm{tr}}\right\rangle^{X^{i}}$ 
denotes the expectation value of the energy-momentum tensor 
$T_{\rho\rho}^{\mathrm{tr}}$ on the light-cone diagram
for the worldsheet 
bosons $X^{i}$ $(i=1,\ldots,24)$ corresponding
to the transverse spacetime coordinates. 

\begin{figure}[htbp]
\begin{center}
\includegraphics[width=0.5\textwidth,keepaspectratio=true]{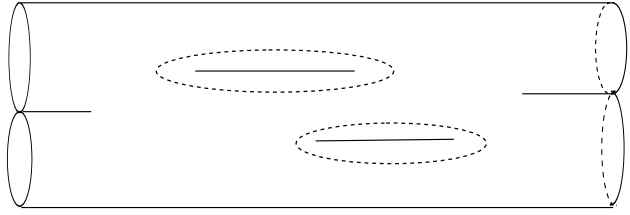}
\end{center}
\caption{The cycles corresponding to the variations 
of the $+$ components of the loop momenta.}
\label{fig:omit}
\end{figure}

What we would like to do in the following is to calculate the 
right hand side of eq.(\ref{eq:deltaZLC1}) and integrate it. 
The variation we consider here corresponds to that of only a subset of 
$6h-6+2N$  moduli parameters. 
We do not consider the variation of $\alpha_\mathcal{I}$'s 
which are not fixed by the momentum conservation, namely 
that of the $+$ components of the loop momenta. 
Such variations correspond to
integration cycles depicted in Figure~\ref{fig:omit}. 
Therefore integrating the right hand side of eq.(\ref{eq:deltaZLC1}), 
integration constants depending on these parameters are left undetermined. 
We will fix these imposing the factorization conditions in subsection 
\ref{sec:factorization-Gamma}. 

\subsection{Integration of the right hand side of eq.(\ref{eq:deltaZLC1})}

\begin{figure}
\begin{centering}
\includegraphics[scale=0.6]{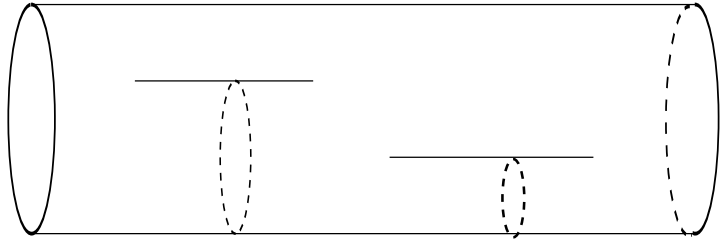}
\par\end{centering}

\caption{The $h$ cycles along which we cut the light-cone diagram
         to make the Riemann surface $\Sigma$ corresponding to
         the light-cone diagram into a surface with no handles
         but with $2h$ holes.
\label{fig:The-Riemann-surface}}
\end{figure}

In order to integrate the right hand side of eq.(\ref{eq:deltaZLC1}), 
we introduce a convenient way to parametrize the moduli of the 
surface. 
As is depicted in Figure \ref{fig:The-Riemann-surface},
cutting along $h$ cycles with constant $\mathop{\mathrm{Re}}\rho$, 
one can make $\Sigma$ into a surface with no handles but with $2h$ holes.
By attaching $2h$ semi-infinite cylinders to the holes, it is possible to
get a tree light-cone diagram, which is denoted by $\tilde{\Sigma}$.
Let $\tilde{\rho}\left(z\right)$ be the Mandelstam mapping which
maps $\mathbb{C}\cup \{\infty\}$ to $\tilde{\Sigma}$:
\begin{eqnarray}
\tilde{\rho}~:~\mathbb{C}\cup\{\infty\}
   &\longrightarrow& \tilde{\Sigma}
\nonumber \\
z \ & \mapsto & \tilde{\rho}\left(z\right)~.
\end{eqnarray}
 $\tilde{\rho}\left(z\right)$ has the form
\begin{equation}
\tilde{\rho}\left(z\right)
 =\sum_{r=1}^{N}\alpha_{r}\ln\left(z-Z_{r}\right)
  +\sum_{A=1}^{h}\beta_{A}\ln\frac{z-Q_{A}}{z-R_{A}}\,.
\label{eq:rhotilde}
\end{equation}
Here $\beta_{A}$ $\left(A=1,\cdots,h\right)$ are 
real positive parameters
corresponding to the lengths of the $h$ cycles 
along which the surface~$\Sigma$ is cut.
The surface $\Sigma$ can be obtained from $\tilde{\Sigma}$ by discarding 
the $2h$ semi-infinite cylinders
and identifying the boundaries. 
Therefore we can use the $z$ coordinate to describe $\Sigma$ 
and we do so in the rest of this subsection. 
The Mandelstam mapping $\rho (z)$ can be given as 
\begin{equation}
\rho (z)=\tilde{\rho}(z)
 + (\mbox{purely imaginary constant}) \,,
\end{equation}
and $\Sigma$ corresponds to $\mathbb{C}\cup \{\infty\}$  with 
disks $D_{Q_{A}},D_{R_{A}}\,\left(A=1,\cdots,h\right)$
around $Q_{A},R_{A}$ excised. 
We identify
$z\in\partial D_{Q_{A}}$ and $w\in\partial D_{R_{A}}$ if 
\begin{equation}
\tilde{\rho}\left(z\right)
 =\tilde{\rho}\left(w\right)+
 i\beta_{A}(\theta_{A}+2\pi n)\,,
\label{eq:thetaA}
\end{equation}
for $n\in\mathbb{Z}$. 

{}From the construction above, one can see that 
it is possible to associate the parameters 
$Z_{r},\beta_{A},\theta_{A},Q_{A},R_{A}$
with any light-cone diagram. Therefore, 
with $\alpha_{r} \, (= 2 p^{+}_{r})$ fixed,
the shape of $\Sigma$ is parametrized locally by 
$Z_{r},\beta_{A},\theta_{A},Q_{A},R_{A}$
modded out by the $6$ conformal transformations on 
$\mathbb{C} \cup \{ \infty \}$.
Thus we have $2N+h+h+2h+2h-6=6h-6+2N$ real parameters, the number
of which coincides with that of the moduli parameters of the punctured
Riemann surface $\Sigma$. 
A variation of the complex structure of $\Sigma$ corresponds to a
variation of these parameters. 
$\beta_A$'s correspond to the loop momenta and 
the variation we consider here corresponds to the one with 
$\delta\beta_A=0$. 
Under such a variation of the parameters, the rule of identification 
(\ref{eq:thetaA}) is also changed as
\begin{equation}
\left(\tilde{\rho}+\delta\tilde{\rho}\right) \left(z\right)
 =\left(\tilde{\rho}+\delta\tilde{\rho}\right)
    \left(w+\delta w\right)
  +i\beta_{A}(\theta_{A}+\delta \theta_A + 2\pi n)\,.
\end{equation}
Accordingly we obtain
\begin{equation}
\delta\tilde{\rho}\left(z\right)
  -\delta\tilde{\rho}\left(w\right)
= \delta w\partial\tilde{\rho}\left(w\right)
+i\beta_{A}\delta \theta_A \,.
\label{eq:delta-thetaA}
\end{equation}
Therefore as a function on $\Sigma$, 
the variation $\delta \rho \left(z\right)$
is discontinuous along the cycle corresponding to $\partial D_{Q_{A}}$. 

In terms of the quantities defined using the $z$ coordinate, 
the right hand side of eq.(\ref{eq:deltaZLC1}) can be expressed as
\begin{equation}
\delta\ln Z^{\mathrm{LC}} 
 =  \sum_{\mathcal{I}}\delta \mathcal{T}_{\mathcal{I}}
      \oint_{C_{\mathcal{I}}}\frac{dz}{2\pi i}
      \frac{1}{\partial\rho (z)}
      \left(\left\langle T_{zz}^{\mathrm{tr}}\right\rangle^{X^{i}}
      -2\{ \rho ,z\}
      \right)
 +\mathrm{c.c.}\,.
\label{eq:deltaZLC}
\end{equation}
Here $\{\rho,z\}$ denotes the Schwarzian derivative,
which is given by
\begin{eqnarray}
-2\left\{ \rho,z\right\} 
&=&
-2\frac{\partial^3\rho}{\partial\rho}
+3\left(\frac{\partial^2\rho}{\partial\rho}\right)^2
\nonumber
\\
&=&
\left(\partial\ln\left|\partial\rho\right|^{2}\right)^{2}
  -2\partial^{2}\ln\left|\partial\rho\right|^{2}\,.
\label{eq:schwarzian}
\end{eqnarray}

\subsubsection*{Calculation of the right hand side of eq.(\ref{eq:deltaZLC})}

$\mathcal{T}_{\mathcal{I}}$ can be expressed as 
\begin{equation}
\mathcal{T}_{\mathcal{I}}
 = \rho\left(z_{\mathcal{I}+}\right)
   - \rho\left(z_{\mathcal{I}-}\right)\,,
\end{equation}
where $z_{\mathcal{I}+}$ and $z_{\mathcal{I}-}$ are the $z$ coordinates
of the interaction points on the two sides of the $\mathcal{I}$-th
internal line. 
Rewriting each term on the right hand side of eq.(\ref{eq:deltaZLC}) as 
\begin{eqnarray}
& &
\delta \mathcal{T}_{\mathcal{I}}
  \oint_{C_{\mathcal{I}}} \frac{dz}{2\pi i}
       \frac{1}{\partial\rho\left(z\right)}
        \left( \left\langle T_{zz}^{\mathrm{tr}}\right\rangle^{X^{i}}
                -2\left\{ \rho,z\right\} 
        \right)
\nonumber \\
& & \qquad \qquad 
   =  \oint_{C_{\mathcal{I}}}\frac{dz}{2\pi i}
       \frac{\delta\rho\left(z\right)
              -\delta\rho\left(z_{\mathcal{I}-}\right)}
            {\partial\rho\left(z\right)}
       \left(\left\langle T_{zz}^{\mathrm{tr}}\right\rangle^{X^{i}} 
               -2\left\{ \rho,z\right\} 
       \right)
\nonumber \\
 &  & \hphantom{\qquad \qquad = \ }
      {}-\oint_{C_{\mathcal{I}}}\frac{dz}{2\pi i}
         \frac{\delta\rho\left(z\right)
                 -\delta\rho\left(z_{\mathcal{I}+}\right)}
              {\partial\rho\left(z\right)}
         \left(
             \left\langle T_{zz}^{\mathrm{tr}}\right\rangle^{X^{i}}
                -2\left\{ \rho,z\right\} 
         \right)\,,
\end{eqnarray}
and deforming the contours, we obtain
\begin{eqnarray}
\delta\ln Z^{\mathrm{LC}} 
& = & -\sum_{r}
      \oint_{Z_{r}}\frac{dz}{2\pi i}
         \frac{\delta\rho\left(z\right)
               -\delta\rho\left(z_{I^{\left(r\right)}}\right)}
               {\partial\rho\left(z\right)}
          \left( \left\langle T_{zz}^{\mathrm{tr}}\right\rangle^{X^{i}}
                  -2\left\{ \rho,z\right\}
          \right)
\nonumber \\
 &  & {}-\sum_{I}
        \oint_{z_{I}}\frac{dz}{2\pi i}
           \frac{\delta\rho\left(z\right)
                  -\delta\rho\left(z_{I}\right)}
                {\partial\rho\left(z\right)}
           \left(\left\langle T_{zz}^{\mathrm{tr}}\right\rangle^{X^{i}}
                  -2\left\{ \rho,z\right\} \right)
\nonumber \\
 &  & {}-\sum_{A}
         \oint_{\partial D_{Q_{A}}}\frac{dz}{2\pi i}
           \frac{\delta\rho\left(z\right)}{\partial\rho\left(z\right)}
           \left(\left\langle T_{zz}^{\mathrm{tr}}\right\rangle^{X^{i}}
                  -2\left\{ \rho,z\right\} \right)
\nonumber \\
 &  & {}-\sum_{A}
          \oint_{\partial D_{R_{A}}}\frac{dz}{2\pi i}
          \frac{\delta\rho\left(z\right)}{\partial\rho\left(z\right)}
           \left(\left\langle T_{zz}^{\mathrm{tr}}\right\rangle^{X^{i}} 
                   -2\left\{ \rho,z\right\} \right)
\nonumber \\
 & & {}+ \mathrm{c.c.}~.
\label{eq:deltaZLC2}
\end{eqnarray}
The third and the fourth terms do not cancel with each other because
of the discontinuity of $\delta\rho\left(z\right)$ mentioned above.

While $\left\langle T_{zz}^{\mathrm{tr}}\right\rangle^{X^{i}} $
is regular for $z\sim Z_{r}$ and $z_{I}$, 
the Schwarzian derivative
$-2\{ \rho,z\}$ given in eq.(\ref{eq:schwarzian})
behaves as 
\begin{eqnarray}
-2\left\{ \rho,z\right\}  
& \sim & 
   \frac{-1}{\left(z-Z_{r}\right)^{2}}
\nonumber \\
& & {}+ \frac{1}{z-Z_{r}}
      \frac{\partial}{\partial Z_{r}}
        \left( 2\sum_{I} G^\mathrm{A}\left(Z_{r};z_{I}\right)
            -2\sum_{s\ne r} G^\mathrm{A}\left(Z_{r};Z_{s}\right)
      {}-\ln g_{Z_{r}\bar{Z}_{r}}^\mathrm{A}
      \right)~~~
\label{eq:rhozZr}
\end{eqnarray}
for $z\sim Z_{r}$ and
\begin{eqnarray}
-2\left\{ \rho,z\right\}  
& \sim & \frac{3}{\left(z-z_{I}\right)^{2}}
\nonumber \\
& & {}+\frac{1}{z-z_{I}}
          \frac{\partial}{\partial z_{I}}
        \left(-2\sum_{J\ne I}
                  G^\mathrm{A}\left(z_{I};z_{J}\right)
              +2\sum_{r} G^\mathrm{A}\left(z_{I};Z_{r} \right)
     {}+3 \ln g_{z_{I}\bar{z}_{I}}^\mathrm{A}
    \right)~~~
\label{eq:rhozzI}
\end{eqnarray}
for $z\sim z_{I}$,
as can be derived using eq.(\ref{eq:metric-z-Arakelov}).  

We here consider a variation of the form
$Z_r\to Z_r+\delta Z_r$, 
$Q_A\to Q_A+\delta Q_A$, 
$R_A\to R_A+\delta R_A$ and 
\begin{eqnarray}
\frac{\delta\rho\left(z\right)
        - \delta \rho \left( z_{I^{\left(r\right)}} \right)}
     {\partial\rho\left(z\right)} 
 & \sim & -\delta Z_{r}
          -\left(z-Z_{r}\right) \delta\bar{N}_{00}^{rr}
           +\mathcal{O}\left(\left(z-Z_{r}\right)^{2}\right)
\qquad \qquad \ \ \mbox{for $z \sim Z_{r}$}~,
\nonumber \\
\frac{\delta\rho\left(z\right)
        -\delta\rho\left(z_{I}\right)}
     {\partial\rho\left(z\right)} 
 & \sim & -\delta z_{I}
          +\left(z-z_{I}\right)
             \frac{1}{2} \delta\left(\ln\partial^{2}\rho\left(z_{I}\right)
                                \right)
          +\mathcal{O}\left(\left(z-z_{I}\right)^{2}\right)
\quad \mbox{for $z \sim z_{I}$}~,
\nonumber \\
\label{eq:delrho-delrho}
\end{eqnarray}
where 
\begin{equation}
\delta z_I
=
z_I(Z_r+\delta Z_r, Q_A+\delta Q_A, R_A+\delta R_A)
-
z_I(Z_r, Q_A, R_A)\,.
\end{equation}
$\bar{N}^{rr}_{00}$ denotes one of the Neumann coefficients and is
given by
\begin{eqnarray}
\bar{N}^{rr}_{00}
 &\equiv& \lim_{z \to Z_{r}}
   \left[ \frac{\rho(z_{I^{(r)}}) - \rho(z)}{\alpha_{r}}
          + \ln (z-Z_{r}) 
   \right]
\nonumber \\
 &=& \frac{\rho (z_{I^{(r)}})}{\alpha_{r}}
     - \sum_{s\neq r} \frac{\alpha_{s}}{\alpha_{r}}
         \ln E(Z_{r},Z_{s})
    + \frac{2\pi i}{\alpha_{r}}
      \int^{Z_{r}}_{P_{0}} \omega
      \frac{1}{\mathop{\mathrm{Im}} \Omega}
          \sum_{s=1}^{N}
                   \alpha_{s} \mathop{\mathrm{Im}} 
                         \int^{Z_{s}}_{P_{0}} \omega~.
\label{eq:Neumann00-holo}
\end{eqnarray}
Using eqs.(\ref{eq:rhozZr}), (\ref{eq:rhozzI}) and (\ref{eq:delrho-delrho}),
we can easily evaluate the first two terms on the right hand side
of (\ref{eq:deltaZLC2}) and obtain 
\begin{eqnarray}
\delta\ln Z^{\mathrm{LC}} 
 & = & \delta
       \left(  -\sum_{r} \bar{N}_{00}^{rr}
               -\sum_{I} \frac{3}{2}
                         \ln\partial^{2}\rho\left(z_{I}\right)
       \right)
\nonumber \\
 &  & {}+\sum_{r} \delta Z_{r}
          \frac{\partial}{\partial Z_{r}}
          \left(2\sum_{I} G^\mathrm{A}\left(Z_{r},z_{I}\right)
                -2\sum_{s\ne r} G^\mathrm{A}\left(Z_{r},Z_{s}\right)
                - \ln g_{Z_{r}\bar{Z}_{r}}^\mathrm{A}
          \right)
\nonumber \\
 &  & {}+\sum_{I} \delta z_{I}
           \frac{\partial}{\partial z_{I}}
           \left( - 2\sum_{J\ne I} G^\mathrm{A}\left(z_{I},z_{J}\right)
                  + 2\sum_{r} G^\mathrm{A}\left(z_{I},Z_{r} \right)
                  + 3 \ln g_{z_{I}\bar{z}_{I}}^\mathrm{A}
           \right)
\nonumber \\
 &  & {}-\sum_{A}
         \oint_{\partial D_{Q_{A}}}
         \frac{dz}{2\pi i}
           \frac{\delta\rho\left(z\right)}{\partial\rho\left(z\right)}
            \left(\left\langle T_{zz}^{\mathrm{tr}}\right\rangle^{X^{i}}
                  -2\left\{ \rho,z\right\} 
            \right)
\nonumber \\
 &  & {}-\sum_{A}
         \oint_{\partial D_{R_{A}}} 
            \frac{dz}{2\pi i}
            \frac{\delta\rho\left(z\right)}{\partial\rho\left(z\right)}
              \left(\left\langle T_{zz}^{\mathrm{tr}}\right\rangle^{X^{i}}
                     -2\left\{ \rho,z\right\}
              \right)
\nonumber \\
& & {}+ \mathrm{c.c.}~.
\label{eq:deltaZLC3}
\end{eqnarray}
In the following, we will show that there exists $Z$ which satisfies
\begin{eqnarray}
\delta\ln Z
 & = & \sum_{r}\delta Z_{r}
       \frac{\partial}{\partial Z_{r}}
       \left( 2\sum_{I} G^\mathrm{A}\left(Z_{r};z_{I}\right)
              -2\sum_{s\ne r} G^\mathrm{A}\left(Z_{r};Z_{s}\right)
              -\ln g_{Z_{r}\bar{Z}_{r}}^\mathrm{A}
      \right)
\nonumber \\
 &  & {} +\sum_{I} \delta z_{I}
      \frac{\partial}{\partial z_{I}}
      \left( -2\sum_{J\ne I} G^\mathrm{A}\left(z_{I};z_{J}\right)
             +2\sum_{r} G^\mathrm{A}\left(z_{I};Z_{r}\right)
             +3 \ln g_{z_{I}\bar{z}_{I}}^\mathrm{A}
      \right)
\nonumber \\
 &  & {}-\sum_{A}\oint_{\partial D_{Q_{A}}}
         \frac{dz}{2\pi i}
         \frac{\delta\rho\left(z\right)}{\partial\rho\left(z\right)}
         \left( \left\langle T_{zz}^{\mathrm{tr}}\right\rangle^{X^{i}}
                -2\left\{ \rho,z\right\}
         \right)
\nonumber \\
 &  & {}-\sum_{A} \oint_{\partial D_{R_{A}}}
         \frac{dz}{2\pi i}
          \frac{\delta\rho\left(z\right)}{\partial\rho\left(z\right)}
          \left( \left\langle T_{zz}^{\mathrm{tr}}\right\rangle^{X^{i}} 
                  -2\left\{ \rho,z\right\} \right)
\nonumber \\
 & & {}+\mathrm{c.c.}\,,
\label{eq:deltalnZ}
\end{eqnarray}
under a variation of the parameters $Z_{r},Q_{A},R_{A}$. 
Then we get 
\begin{equation}
\delta\ln Z^{\mathrm{LC}} 
  =  \delta
       \left(  -\sum_{r} 2 \mathop{\mathrm{Re}} 
                            \bar{N}_{00}^{rr}
               -\sum_{I} \frac{3}{2} \ln |\partial^{2}\rho (z_{I})|^{2}
              +\ln Z
       \right)\,.
\label{eq:deltalnZLC-Z}
\end{equation}

\subsubsection*{Correlation function $Z$}

Let us consider a metric
$ds^2=2g_{z\bar{z}}dzd\bar{z}$
on the Riemann surface $\Sigma$ and define 
\begin{eqnarray}
Z & \equiv &
     Z^{X} [g_{z\bar{z}}]^{23}
     \int_{\Sigma} \left[d\Phi\right]_{g_{z\bar{z}}}
       e^{-S\left[\Phi\right]}
       \delta \left( \Phi\left(z_{0},\bar{z}_{0} \right)
              \right)
       \prod_{I=1}^{2h-2+N} \mathcal{O}_{I}
       \prod_{r=1}^{N}V_{r}\,.
\label{eq:Z}
\end{eqnarray}
Here $S\left[\Phi\right]$ is the action
for the boson $\Phi$ given as
\begin{equation}
S\left[\Phi\right]
  \equiv
  \frac{1}{8\pi}
  \int d^{2}z\sqrt{g}
  \left[g^{ab} \partial_{a}\Phi\partial_{b}\Phi
        - 2\sqrt{2}iR\Phi\right]\,,
\label{eq:actionPhi}
\end{equation}
and $Z^{X}\left[g_{z\bar{z}}\right]$ is given in eq.(\ref{eq:def-ZX}). 
$\mathcal{O}_{I}$ and $V_{r}$ are the vertex
operators defined as
\begin{eqnarray}
\mathcal{O}_{I} 
  & \equiv & 
      \left(2 g_{z_{I}\bar{z}_{I}}\right)^{2}
       e^{i\sqrt{2}\Phi}\left(z_{I},\bar{z}_{I}\right)\,,
  \nonumber \\
V_{r} 
  & \equiv &
  \left( 2 g_{Z_{r}\bar{Z}_{r}} \right)^{-2}
  e^{-i\sqrt{2}\Phi}\left(Z_{r},\bar{Z}_{r}\right)\,.
\label{eq:Vr}
\end{eqnarray}
The operators 
$e^{i\sqrt{2}\Phi}\left(z_{I},\bar{z}_{I}\right),
\, e^{-i\sqrt{2}\Phi}\left(Z_{r},\bar{Z}_{r}\right)$
on the right hand side 
are normal ordered and $\mathcal{O}_{I} , V_{r}$ are
defined to be Weyl invariant.
$\delta\left(\Phi\left(z_{0},\bar{z}_{0}\right)\right)$
is necessary to soak up the zero mode of $\Phi$
and $Z$ does not
depend on $z_{0},\bar{z}_{0}$. 
The energy-momentum tensor for $\Phi$ is given as 
\begin{equation}
-\frac{1}{2}\partial\Phi\partial\Phi
 -\sqrt{2}i
 \left(\partial-\partial\ln g_{z\bar{z}}\right) \partial\Phi\,,
\end{equation}
and the Virasoro central charge is $-23$. 
Since $\mathcal{O}_{I},V_{r}$
are Weyl invariant, $Z$ does not depend on the metric $g_{z\bar{z}}$
and is a function of the moduli parameters 
$Z_{r},\beta_{A},\theta_{A},Q_{A},R_{A}$.
With the metric $g_{z\bar{z}}$ taken to be the Arakelov metric 
$g_{z\bar{z}}^\mathrm{A}$, 
we can evaluate $Z$ to be
\begin{eqnarray}
Z
 & \propto& Z^{X} [g_{zz}^\mathrm{A} ]^{24}
            \prod_{I}\left(2g_{z_{I}\bar{z}_{I}}^\mathrm{A}\right)^{3}
            \prod_{r}\left(2g_{Z_{r}\bar{Z}_{r}}^\mathrm{A}\right)^{-1}
\nonumber \\
 &  & \quad \times
      \exp\left[-2\sum_{I<J}G^\mathrm{A}\left(z_{I},z_{J}\right)
           -2\sum_{r<s}G^\mathrm{A}\left(Z_{r};Z_{s}\right)
           +2\sum_{I,r}G^\mathrm{A}\left(z_{I};Z_{r}\right)\right]\,,
\label{eq:Zexplicit}
\end{eqnarray}
using the Arakelov Green's function $G^\mathrm{A}(z;w)$
with respect to the Arakelov metric. 
We note that unlike the scalar field used in bosonization,
$\Phi$ is not circle valued.
Therefore contributions from the soliton sector are not
included in eq.(\ref{eq:Zexplicit}).

In the following, we would like to prove that $Z$ thus defined  
satisfies eq.(\ref{eq:deltalnZ}). 
Let us first rewrite $Z$ in a factorized form. 
As we did earlier, 
cutting the light-cone
diagram $\Sigma$ and attaching semi-infinite cylinders, one can get
a tree light-cone diagram $\tilde{\Sigma}$. We replace the cut propagators by 
\begin{equation}
\sum_{n}e^{-Th_{n}}\left|n\right\rangle \left\langle n\right|
=
\sum_{n}e^{-Th_{n}}
\mathcal{O}_n (0,0)
\left|0\right\rangle \left\langle 0\right|
 I \circ \mathcal{O}_n (0,0)
\,,
\end{equation}
where $\{ |n\rangle \}$ is a complete basis of the states, 
$h_{n}$ denotes the weight of $|n\rangle$,
$\mathcal{O}_n(z,\bar{z})$ is the local operator corresponding 
to the state $|n\rangle$
and $I(z) \equiv 1/z$ is the inversion.
We denote by $f \circ \mathcal{O} (z,\bar{z})$
the transform of the local operator $\mathcal{O} (z,\bar{z})$
under the conformal transformation
$(z,\bar{z}) \to (f(z),\bar{f} (\bar{z}))$.
One can express $Z$ in terms of a correlation function on 
$\mathbb{C}\cup\{\infty\}$ as
\begin{eqnarray}
Z & = & \int_{\mathbb{C}\cup\{\infty\}}
        \left[dX^{1}\cdots dX^{23}d\Phi \right]
         e^{-S\left[\Phi\right]-S^{X}}
        \,\delta\left(\Phi\left(z_{0},\bar{z}_{0}\right)\right)
        \prod_{I}\mathcal{O}_{I}
        \prod_{r}V_{r}
\nonumber \\
 &  & \hphantom{
                \int_{\mathbb{C}}
                \left[dX^{1}\cdots dX^{23}d\Phi\right]
                 e^{-S\left[\Phi\right]} \,
                }
       \times \prod_{A} 
          \left(\sum_{n}
               f_{Q_{A}}^{-1}\circ\mathcal{O}_{n}\left(0,0\right)
               f_{R_{A}}^{-1}\circ\mathcal{O}_{n}\left(0,0\right)
         \right),~~~~
\label{eq:ZmathbbC}
\end{eqnarray}
where $S^{X}$ denotes the worldsheet action 
for the free bosons $X^{1},\ldots,X^{23}$, and
\begin{equation}
f_{Q_{A}}\left(z\right) 
 =  e^{\frac{1}{\beta_{A}}\tilde{\rho}\left(z\right)}\,,
\qquad 
f_{R_{A}}\left(z\right) 
  = e^{-\frac{1}{\beta_{A}}\tilde{\rho}\left(z\right)}\,.
\label{eq:XPhi}
\end{equation}
Since the total central charge of the system vanishes, 
we do not have to specify the metric on $\mathbb{C}\cup\{\infty\}$ in 
eq.(\ref{eq:ZmathbbC}). 

For any $f\left(z\right)$ regular at $z=0$, 
there exists an operator
$U_{f}$ of the form~\cite{LeClair:1988sj,Sen:1989gf,
Rastelli:2000iu,Schnabl:2002gg}
\begin{equation}
U_{f}=e^{\oint\frac{dz}{2\pi i}v\left(z\right)T_{zz}
         + \mathrm{c.c.}}\,,
\end{equation}
such that 
\begin{equation}
U_{f}\mathcal{O}\left(0,0\right)|0\rangle 
=
f\circ\mathcal{O}\left(0,0\right)|0\rangle 
\,.
\end{equation}
The relation between $v\left(z\right)$ and $f\left(z\right)$ is
given by
\begin{equation}
e^{v\left(z\right)\partial_{z}}z=f\left(z\right)\,,
\label{eq:vf}
\end{equation}
and one can get $v(z)$ from $f(z)$ solving eq.(\ref{eq:vf}). 
Suppose $f\to f+\delta f$ is an infinitesimal variation of $f$. 
From 
\begin{equation}
\left(1+\frac{\delta f}{\partial f}\partial_z\right)
e^{v\left(z\right)\partial_{z}}z
=
f\left(z\right)+\delta f(z)\,,
\end{equation}
one can prove 
\begin{equation}
U_{f}\left( 1 + 
           \left(\oint\frac{dz}{2\pi i}
                  \frac{\delta f}{\partial f} T_{zz}
                + \mathrm{c.c.}
           \right)
      \right)
=U_{f+\delta f}\,.\label{eq:deltaf}
\end{equation}
Using eq.(\ref{eq:deltaf}), $\delta\ln Z$ under the variation of 
$Z_{r},Q_{A},R_{A}$
is given as 
\begin{eqnarray}
\delta\ln Z
 &=& \frac{1}{Z}
     \int_{\mathbb{C} \cup \{ \infty \}}
          \left[dX^{1}\cdots dX^{23}d\Phi\right]
     e^{-S\left[\Phi\right]-S^{X}}
     \delta\left(\Phi\left(z_{0},\bar{z}_{0}\right)\right)
\nonumber \\
 &  & \hphantom{\frac{1}{Z}\int 
                }
      \times\left[ 
         \sum_{I}\delta z_{I}\partial\mathcal{O}_{I}
             \prod_{J\ne I}\mathcal{O}_{J}
             \prod_{r}V_{r}\prod_{A}
               \left(\sum_{n}
                       f_{Q_{A}}^{-1}\circ\mathcal{O}_{n}
                         \left(0,0\right)
                       f_{R_{A}}^{-1}\circ\mathcal{O}_{n}
                         \left(0,0\right)
              \right)\right.
\nonumber \\
 &  & \hphantom{
                \frac{1}{Z}  \int
                  \times
                }
    \quad
    {} + \sum_{r}\delta Z_{r}\partial V_{r}
           \prod_{I}\mathcal{O}_{I}
           \prod_{s\ne r}V_{s}
           \prod_{A}
             \left( \sum_{n} f_{Q_{A}}^{-1}\circ\mathcal{O}_{n}
                               \left(0,0\right)
                             f_{R_{A}}^{-1}\circ\mathcal{O}_{n}
                                \left(0,0\right)
             \right)
\nonumber \\
 &  & \hphantom{
                \frac{1}{Z}\int
                 \times
               }
     \quad
      {} +\prod_{I}\mathcal{O}_{I}\prod_{r}V_{r}
           \sum_{A} 
             \left( -\oint_{Q_{A}}\frac{dz}{2\pi i}
                     \frac{\delta\tilde{\rho}}{\partial\tilde{\rho}}
                       T_{zz}
                    -\oint_{R_{A}}\frac{dz}{2\pi i}
                      \frac{\delta\tilde{\rho}}{\partial\tilde{\rho}}
                       T_{zz}
             \right)
\nonumber \\
 &  & \hphantom{
               \frac{1}{Z}\int
                \times\quad+\prod_{I}\mathcal{O}_{I}\prod_{r}V_{r}
               }
      \times  \prod_{B}
       \left( \sum_{n}f_{Q_{B}}^{-1}\circ\mathcal{O}_{n}
                        \left(0,0\right)
                      f_{R_{B}}^{-1}\circ\mathcal{O}_{n}
                        \left(0,0\right)
      \right)
\nonumber \\
 &  & \hphantom{
               \frac{1}{Z}\int
                \times \left[ \right.
               }
 \left.
       \vphantom{
               \sum_{I}\delta z_{I}\partial\mathcal{O}_{I}
                 \prod_{J\ne I}\mathcal{O}_{J}
                 \prod_{r}V_{r}\prod_{A}
                   \left(\sum_{n}
                          f_{Q_{A}}^{-1}\circ\mathcal{O}_{n}
                             \left(0,0\right)
                          f_{R_{A}}^{-1}\circ\mathcal{O}_{n}
                             \left(0,0\right)\right)
                }
    {}+\mathrm{c.c.}\right]\,.
\label{eq:deltalnZ2}
\end{eqnarray}
The right hand side of eq.(\ref{eq:deltalnZ2}) can be written in terms
of the correlation functions on $\Sigma$:
\begin{eqnarray}
\delta\ln Z
 &=& \frac{1}{Z} \int_{\Sigma}
     \left[dX^{1}\cdots dX^{23}d\Phi\right]
      e^{-S\left[\Phi\right]-S^{X}} 
      \delta\left(\Phi\left(z_{0},\bar{z}_{0}\right)\right)
 \nonumber \\
 &  & \hphantom{
                \frac{1}{Z}\int_{\mathbb{C}}
               }
\times\left[\sum_{I}\delta z_{I}\partial\mathcal{O}_{I}
               \prod_{J\ne I}\mathcal{O}_{J}
               \prod_{r}V_{r}
             +\sum_{r}\delta Z_{r}\partial V_{r}
                \prod_{I}\mathcal{O}_{I}
                \prod_{s\ne r}V_{s} 
        \right.
 \nonumber \\
 &  & \hphantom{
                \frac{1}{Z}\int_{\mathbb{C}}
                \quad\times
               }
  \quad {}+\prod_{I}\mathcal{O}_{I}
           \prod_{r}V_{r}
             \sum_{A} 
                \left(-\oint_{Q_{A}}\frac{dz}{2\pi i}
                       \frac{\delta \rho}
                              {\partial \rho}
                       T_{zz}
                     -\oint_{R_{A}}\frac{dz}{2\pi i}
                       \frac{\delta \rho}
                              {\partial \rho}
                       T_{zz}
                \right)
\nonumber \\
 &  & \hphantom{
                \frac{1}{Z}\int_{\mathbb{C}}
               }
       \left.
         \vphantom{
                   \sum_{I}\delta z_{I}\partial\mathcal{O}_{I}
                     \prod_{J\ne I}\mathcal{O}_{J}
                     \prod_{r}V_{r}\prod_{A}
                        \left(\sum_{n}f_{Q_{A}}^{-1}\circ\mathcal{O}_{n}
                                       \left(0,0\right)
                                      f_{R_{A}}^{-1}\circ\mathcal{O}_{n}
                                       \left(0,0\right)
                        \right)
                   }
 \qquad {} +\mathrm{c.c.}\right]\,.
\label{eq:ZSigma}
\end{eqnarray}
It is straightforward to evaluate 
the contribution of the first two terms of the parenthesis 
and we get the first two terms on the right hand side of 
eq.(\ref{eq:deltalnZ}).
The energy-momentum tensor $T_{zz}$ is given as
\begin{equation}
T_{zz}
= - \frac{1}{2}\left(\partial\Phi\right)^{2}
  -\sqrt{2}i\left(\partial-\partial\ln g_{z\bar{z}}^\mathrm{A}\right)
      \partial\Phi
  -\sum_{i=1}^{23}\frac{1}{2}\left(\partial X^{i}\right)^{2}
  + \left(\partial\ln g_{z\bar{z}}^\mathrm{A}\right)^{2}
  - 2\partial^{2}\ln g_{z\bar{z}}^\mathrm{A}~,
\label{eq:EM-PhiX}
\end{equation}
if we take the metric on $\Sigma$ to be the Arakelov metric. 
The expectation value of $T_{zz}$ can be calculated to be
\begin{eqnarray}
\left\langle T_{zz}\right\rangle  
&\equiv& 
  \frac{1}{Z} \int_{\Sigma} \left[ dX^{1} \cdots dX^{23} d\Phi \right]
  e^{-S \left[ \Phi \right] -S^{X}}
  \, T_{zz}
  \,\delta \left( \Phi (z_{0},\bar{z}_{0}) \right)
  \prod_{I} \mathcal{O}_{I} \prod_{r} V_{r}
\nonumber \\
& = & {}-\frac{1}{2}\left(\partial\Phi_{\mathrm{cl}}\right)^{2}
      -\sqrt{2}i\left(\partial-\partial\ln g_{z\bar{z}}^\mathrm{A}\right)
       \partial\Phi_{\mathrm{cl}}
\nonumber \\
 &  & {}+ 24\lim_{w\to z}
      \left(  -\frac{1}{2}\partial_{z}\partial_{w}G^\mathrm{A}\left(z;w\right)
              -\frac{\frac{1}{2}}{\left(z-w\right)^{2}}
      \right)
      +\left(\partial\ln g_{z\bar{z}}^\mathrm{A}\right)^{2}
      -2\partial^{2}\ln g_{z\bar{z}}^\mathrm{A}
\nonumber \\
 & = & \left\langle T_{zz}^{\mathrm{tr}} \right\rangle^{X^{i}} 
       -2\left\{ \rho,z\right\} \,,
\label{eq:expectationT}
\end{eqnarray}
where
\begin{equation}
\Phi_{\mathrm{cl}}\left(z,\bar{z}\right)
 =i \sqrt{2} \sum_{I}G^\mathrm{A}\left(z;z_{I}\right)
  -i \sqrt{2} \sum_{r}G^\mathrm{A}\left(z;Z_{r}\right)+c'\,,
\label{eq:Phicl}
\end{equation}
and $c'$ is a constant which is fixed by the condition 
$\Phi_{\mathrm{cl}}\left(z_{0},\bar{z}_{0}\right)=0$.
Substituting eq.(\ref{eq:expectationT}) into eq.(\ref{eq:ZSigma}), 
we get eq.(\ref{eq:deltalnZ}) and thus eq.(\ref{eq:deltalnZLC-Z}).

\subsubsection*{Partition function $Z^\mathrm{LC}$}

Substituting 
eq.(\ref{eq:Zexplicit})
into eq.(\ref{eq:deltalnZLC-Z}), 
we eventually obtain
\begin{equation}
Z^{\mathrm{LC}} = Z^{X} [g_{z\bar{z}}^\mathrm{A}]^{24}
                  e^{-\Gamma 
                       \left[g_{z\bar{z}}^\mathrm{A},\,\ln |\partial \rho|^{2}
                       \right] }~,
\label{eq:ZLC-critical}
\end{equation}
where 
\begin{eqnarray}
e^{-\Gamma \left[ g_{z\bar{z}}^\mathrm{A},\,\ln |\partial \rho|^{2}
           \right]}
& = & \mathcal{C}(\beta_A )
     \prod_{r} \left[ e^{-2\mathrm{Re}\bar{N}_{00}^{rr}}
                      \left( 2 g_{Z_{r}\bar{Z}_{r}}^\mathrm{A}\right)^{-1}
                \right]
     \prod_{I} \left[ \left|\partial^{2} \rho \left(z_{I}\right)
                      \right|^{-3}
                      \left(2g_{z_{I}\bar{z}_{I}}^\mathrm{A}\right)^{3}
               \right]
\nonumber \\
 &  & \times
       \exp\left[-2\sum_{I<J}G^\mathrm{A}\left(z_{I};z_{J}\right)
             -2\sum_{r<s}G^\mathrm{A}\left(Z_{r};Z_{s}\right)
             +2\sum_{I,r}G^\mathrm{A}\left(z_{I};Z_{r}\right)\right].
\nonumber \\
\label{eq:Gamma}
\end{eqnarray}
Here $\mathcal{C}(\beta_A )$ is an 
integration constant independent of the parameters 
$Z_{r},Q_{A},R_{A}$. 
Using eq.(\ref{eq:partial2rhozI}), one can express 
$e^{-\Gamma \left[ g_{z\bar{z}}^\mathrm{A},\,\ln |\partial \rho|^{2}
            \right]}$
given in eq.(\ref{eq:Gamma}) as
\begin{equation}
e^{-\Gamma \left[ g_{z\bar{z}}^\mathrm{A},\,\ln |\partial \rho|^{2} \right]}
 = \mathcal{C}(\beta_A ) e^{-2(h-1)c}
   \prod_{r} \left[ e^{-2\mathop{\mathrm{Re}} \bar{N}^{rr}_{00}}
                    (2g_{Z_{r}\bar{Z}_{r}}^\mathrm{A})^{-1}
                    \alpha_{r}^{-2} \right]
   \prod_{I} \left[ \left| \partial^{2} \rho (z_{I}) \right|^{-1}
                    2g_{z_{I}\bar{z}_{I}}^\mathrm{A} \right]~.
\label{eq:expressionI-Gamma}
\end{equation}

Applying the bosonization 
technique~\cite{AlvarezGaume:1987vm,Verlinde:1986kw,
Dugan:1987qe,Sonoda:1987ra,D'Hoker:1989ae}
to the $bc$ system in which the weight
of the $b$-ghost is $1$ and that of the $c$-ghost is
$0$,
one can have the following expression of 
$Z^{X}[g^{\mathrm{A}}_{z\bar{z}}]$,
\begin{equation}
Z^{X}[g^{\mathrm{A}}_{z\bar{z}}]^{24}
\propto  e^{2\delta (\Sigma)}~,
\label{eq:ZX-delta}
\end{equation}
in terms of the Faltings' invariant
$\delta (\Sigma)$~\cite{Faltings:1984}
defined by
\begin{eqnarray}
e^{ -\frac{1}{4} \delta (\Sigma)}
&=& \left(\det \mathop{\mathrm{Im}} \Omega \right)^{\frac{3}{2}}
  \left| \theta [\xi] (0|\Omega) \right|^{2}
      \frac{\prod_{i=1}^{h} 
              \left(2g^{\mathrm{A}}_{\hat{z}_{i}\bar{\hat{z}}_{i}} 
              \right)}
           {\left| \det \omega_{j}(\hat{z}_{i}) \right|^{2}}
\nonumber \\
&& \ \times
      \exp \left[ -\sum_{i<j} G^{\mathrm{A}} (\hat{z}_{i};\hat{z}_{j})
                  + \sum_{i} G^{\mathrm{A}} (\hat{z}_{i};\hat{w}) \right].
\label{eq:Faltings}
\end{eqnarray}
Here $\hat{z}_{i}$ $(i=1,\ldots,h)$ and $\hat{w}$ are arbitrary points
on $\Sigma$, and $\xi \in J (\Sigma)$ is defined as
\begin{equation}
\xi \equiv \sum_{i=1}^{h} \int^{\hat{z}_{i}}_{P_{0}} \omega
        - \int^{\hat{w}}_{P_{0}} \omega 
        - \Delta~.
\label{eq:zeta}
\end{equation}
Putting eqs.(\ref{eq:expressionI-Gamma}) and (\ref{eq:ZX-delta})
together, we obtain the following expression of $Z^{\mathrm{LC}}$,
\begin{equation}
Z^{\mathrm{LC}}
 = \mathcal{C}_{h,N}(\beta_A ) e^{-2(h-1)c} e^{2\delta (\Sigma)}
   \prod_{r} \left[ e^{-2\mathop{\mathrm{Re}} \bar{N}^{rr}_{00}}
                    (2g^{\mathrm{A}}_{Z_{r}\bar{Z}_{r}})^{-1}
                    \alpha_{r}^{-2} \right]
   \prod_{I} \left[ \left| \partial^{2} \rho (z_{I}) \right|^{-1}
                    2g^{\mathrm{A}}_{z_{I}\bar{z}_{I}} \right].
\label{eq:Z-LC-Arakelov}
\end{equation}
The factor $\mathcal{C}_{h,N}(\beta_A )$ is left undetermined. 
Since the expression~(\ref{eq:Z-LC-Arakelov}) is given in terms of 
the quantities which is independent of the choice of the local complex 
coordinate $z$,
 eq.(\ref{eq:Z-LC-Arakelov}) is valid 
for the coordinates other than 
the  $z$  coordinate used in this subsection to derive it.

\subsection{Factorization}
\label{sec:factorization-Gamma}

We can fix the $\mathcal{C}_{h,N}(\beta_A)$ 
using the factorization condition. 
By varying lengths of the propagators, 
it is possible to realize the degeneration limit of 
the Riemann surface~$\Sigma$.
Taking such a limit of $Z^{\mathrm{LC}}$ in eq.(\ref{eq:Z-LC-Arakelov}) 
and imposing the factorization conditions, we are able to 
get relations which $\mathcal{C}_{h,N}(\beta_A)$'s with various 
$h,N$ should satisfy. 
In the following, we first consider the degeneration
of an $h$-loop diagram with $h>1$ depicted in 
Figure~\ref{fig:plumbing} and show that $\mathcal{C}_{h,N}(\beta_A)$ 
with $h>1$ can be expressed by $\mathcal{C}_{1,N}(\beta_A)$. 
We then consider the degeneration depicted in Figure~\ref{fig:h=1}, in which
a one-loop diagram is separated into two tree diagrams. 
The partition functions for the tree diagrams are given in 
Ref.~\cite{Baba:2009ns} and 
we are able to get $\mathcal{C}_{1,N}(\beta_A)$.

\subsubsection{$h>1$ case}

\begin{figure}[htb]

\begin{center}
\includegraphics[width=0.95\textwidth]{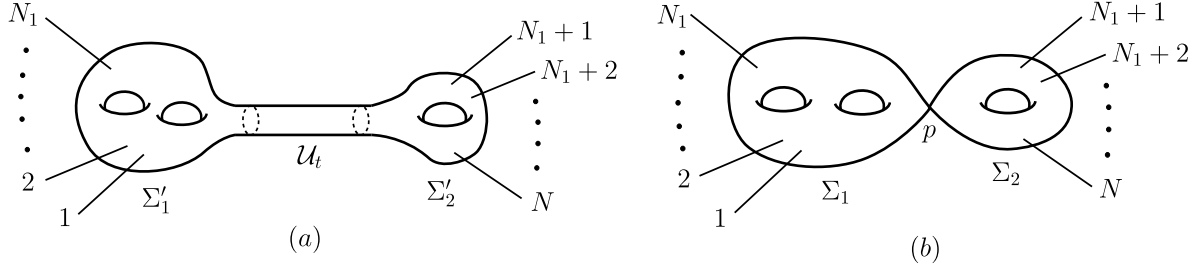}
\end{center}

\caption{The degeneration of Riemann surface $\Sigma$.
        (a) The degeneration process described by the
           plumbing fixture $\mathcal{U}_{t}$.
        (b) A cycle homologous to zero is pinched to the point $p$.}
\label{fig:plumbing}

\end{figure}

In the degeneration depicted in Figure~\ref{fig:plumbing}, 
a zero homology cycle is pinched to
a node $p$ and the worldsheet Riemann surface $\Sigma$
will be separated into two disconnected components
$\Sigma_{1}$ and $\Sigma_{2}$ in Figure~\ref{fig:plumbing}~(b).
We denote the node $p$ by $p_{1}$ or $p_{2}$ depending on
whether it is regarded as
a puncture in $\Sigma_{1}$ or that in $\Sigma_{2}$.
Let $\Sigma_{1}$, $\Sigma_{2}$ be of genus $h_{1}$, $h_{2}$
and with $N_{1}+1$, $N_{2}+1$ punctures respectively,
where $h_{1}+h_{2}=h$ and $N_{1}+N_{2}=N$.
Since the partition functions on $\Sigma_1$
and $\Sigma_2$ 
are again expressed as eq.(\ref{eq:Z-LC-Arakelov}), we can obtain 
a relation between $\mathcal{C}_{h,N}(\beta )$ and 
$\mathcal{C}_{h_1,N_1}(\beta_{A_1})\mathcal{C}_{h_2,N_2}(\beta_{A_2})$ 
by examining such degenerations.

The degeneration that we consider here
can be described as the process in which
the plumbing fixture $\mathcal{U}_{t}$ parametrized by a complex
parameter $t$ with $|t|<1$ disconnects the Riemann surface
$\Sigma (t)$
into $\Sigma_{1}$ and $\Sigma_{2}$ as $t \to 0$.
In Figure~\ref{fig:plumbing}~(a), we denote by
$\Sigma'_{1}$ and $\Sigma'_{2}$ the components of
the complement of $\mathcal{U}_{t}$ in $\Sigma (t)$.
In this process, the basis of the homology cycles
$\{a_{j},b_{j}\}$ $(j=1,\ldots,h)$ will be
divided into the homology basis on $\Sigma_{1}$
and that on $\Sigma_{2}$.
We assume that $\{a_{j},b_{j}\}$
are ordered so that
$\{a_{j_{1}}, b_{j_{1}}\}$ $(1 \leq j_{1} \leq h_{1})$ are
cycles in $\Sigma'_{1}$
and $\{a_{j_{2}},b_{j_{2}}\}$ $(h_{1}+1 \leq j_{2} \leq h)$
are those in $\Sigma'_{2}$.
Similarly, the set of the punctures 
$Z_{r}$~$(r=1,\ldots,N)$ 
will be separated as $Z_{r}=(Z_{r_{1}},Z_{r_{2}})$
($1\leq r_{1} \leq N_{1}$, $N_{1}+1 \leq r_{2} \leq N$)
with $Z_{r_{1}} \in \Sigma'_{1}$ and $Z_{r_{2}} \in \Sigma'_{2}$,
and the set of the interaction points 
$z_{I}$~$(I=1,\ldots,2h-2+N)$
as $z_{I}=(z_{I_{1}},z_{I_{2}})$
($1\leq I_{1} \leq 2h_{1}-1+N_{1}$,
  $2h_{1}+N_{1} \leq I_{2} \leq 2h-2 +N$)
with $z_{I_{1}} \in \Sigma'_{1}$ and $z_{I_{2}} \in \Sigma'_{2}$.
Without the loss of generality,
we assume that the base point $P_{0}$ of the Abel-Jacobi map
on $\Sigma$ lies in $\Sigma'_{1}$.

\subsubsection*{Asymptotics of Arakelov Green's function and Arakelov metric}

The canonical basis of the holomorphic one-forms
$\omega_{j} (z;t)$~$(j=1,\ldots,h)$ of $\Sigma (t)$
tends to the combined bases of 
holomorphic one-forms $\omega^{(1)}_{j_{1}}(z)$
$(1\leq j_{1} \leq h_{1})$ and
$\omega^{(2)}_{j_{2}} (z)$ $(h_{1}+1\leq j_{2} \leq h)$
of $\Sigma_{1}$ and $\Sigma_{2}$ 
as~\cite{Fay1973,Wentworth:1991,Yamada:1980}
\begin{equation}
\omega_{j_{1}} (z;t)
 = \left\{ 
   \begin{array}{ll}
     \displaystyle
     \omega^{(1)}_{j_{1}} (z)
        + \mathcal{O} (t^{2})
      & \mbox{for $z \in \Sigma'_{1}$}\\[2ex]
     \displaystyle
      -t \omega^{(1)}_{j_{1}} (p_{1}) \omega^{(2)} (z,p_{2})
       + \mathcal{O} (t^{2})
       & \mbox{for $z \in \Sigma'_{2}$}
    \end{array}
   \right.~,
\label{eq:asymptotic-one-form}
\end{equation}
and similarly for $\omega_{j_{2}} (z;t)$
with the roles of $\Sigma_{1}$ and $\Sigma_{2}$
interchanged.
Here 
$\omega^{(1)} (z,w)$ denotes
the abelian differential
of the second kind on $\Sigma_{1}$ defined as
\begin{equation}
\omega^{(1)} (z,w)
 = dz \frac{\partial^{2}}{\partial z \partial w}
       \ln E_{1} (z,w)~,
\end{equation}
and similarly for $\omega^{(2)}(z,w)$,
where $E_{1}(z,w)$, $E_{2}(z,w)$ denote the prime forms
on $\Sigma_{1}$, $\Sigma_{2}$ respectively.
Integrating $\omega_{j} (z;t)$ in eq.(\ref{eq:asymptotic-one-form})
over the $b$ cycles, we obtain the behavior 
of the period matrix $\Omega (t)$ of $\Sigma (t)$,
\begin{equation}
\Omega (t)
 = \left( \begin{array}{cc}
            \left(\Omega_{1}\right)_{i_{1}j_{1}} & 0 \\
             0 & \left(\Omega_{2}\right)_{i_{2}j_{2}}
          \end{array}
   \right)
    - i 2\pi t
     \left( 
       \begin{array}{cc}
         0
       & \omega^{(1)}_{i_{1}}(p_{1}) \omega_{j_{2}}^{(2)} (p_{2})\\
         \omega^{(2)}_{i_{2}}(p_{2}) \omega_{j_{1}}^{(1)} (p_{1})
       & 0
       \end{array}
   \right)
   + \mathcal{O}(t^{2})~,
\label{eq:asymptotic-period}
\end{equation}
where $\Omega_{1}$ and $\Omega_{2}$ denote
the period matrices of $\Sigma_{1}$ and $\Sigma_{2}$.
Substituting eqs.(\ref{eq:asymptotic-one-form})
and (\ref{eq:asymptotic-period}) into the definition~(\ref{eq:primeform})
of the prime form yields
\begin{eqnarray}
E (z_{1},w_{1} ) &\sim& E_{1} (z_{1},w_{1})
\hspace{9em} \mbox{for $z_{1},w_{1} \in \Sigma'_{1}$}~,
\nonumber \\
E (z_{2},w_{2}) &\sim& E_{2} (z_{2},w_{2})
\hspace{9em} \mbox{for $z_{2},w_{2} \in \Sigma_{2}$}~,
\nonumber \\
E(z_{1},z_{2}) &\sim& E_{1} (z_{1},p_{1})E_{2}(p_{2},z_{2}) (-t)^{-\frac{1}{2}}
\qquad \mbox{for $z_{1} \in \Sigma'_{1}$,\ \ $z_{2} \in \Sigma'_{2}$}~.
\label{eq:asymptotic-primeform}
\end{eqnarray}
Plugging eq.(\ref{eq:asymptotic-one-form}) into 
eq.(\ref{eq:Riemannconst}), we have
\begin{equation}
\Delta_{j_{1}} 
\sim \Delta^{(1)}_{j_{1}}
   + h_{2} \int^{p_{1}}_{P_{0}} \omega^{(1)}_{j_{1}}~,
\qquad
\Delta_{j_{2}}
 \sim
  \Delta^{(2)}_{j_{2}}
 - (h_{2}-1) \int^{p_{2}}_{P'_{0}} \omega^{(2)}_{j_{2}}~,
\label{eq:asymptotic-Delta}
\end{equation}
where $\Delta^{(1)}$, $\Delta^{(2)}$ denote
the vectors of Riemann constants
of $\Sigma_{1}$, $\Sigma_{2}$ 
for the base points $P_{0}$, $P'_{0}$ respectively.
Here 
$P'_{0}$ is an arbitrary point on $\Sigma'_{2}$
and we will take it as the base point of the Abel-Jacobi map
on $\Sigma_{2}$ throughout the subsequent analyses.

It is proved in Ref.~\cite{Wentworth:1991} that
the Arakelov Green's function $G^{\mathrm{A}} (z;w)$
on the degenerating surface $\Sigma$ behaves as
\begin{eqnarray}
G^{\mathrm{A}} (z_{1},w_{1}) 
 &\sim& -2 \left( \frac{h_{2}}{h} \right)^{2} \ln |\tau|
        + G^{\mathrm{A}}_{1} (z_{1},w_{1})
        - \frac{h_{2}}{h} G^{\mathrm{A}}_{1} (z_{1},p_{1})
        - \frac{h_{2}}{h} G^{\mathrm{A}}_{1} (w_{1},p_{1})~,
\nonumber \\
G^{\mathrm{A}} (z_{2},w_{2})
 &\sim& -2 \left( \frac{h_{1}}{h} \right)^{2} \ln |\tau|
        + G^{\mathrm{A}}_{2} (z_{2},w_{2})
        - \frac{h_{1}}{h} G^{\mathrm{A}}_{2} (z_{2},p_{2})
        - \frac{h_{1}}{h} G^{\mathrm{A}}_{2} (w_{2},p_{2})~,
\nonumber \\
G^{\mathrm{A}} (z_{1},z_{2})
 &\sim& 2 \frac{h_{1}h_{2}}{h^{2}} \ln |\tau|
        + \frac{h_{1}}{h} G^{\mathrm{A}}_{1} (z_{1},p_{1})
        + \frac{h_{2}}{h} G^{\mathrm{A}}_{2} (z_{2},p_{2})~,
\label{eq:G-t}
\end{eqnarray}
for $z_{1},w_{1} \in \Sigma'_{1}$ and $z_{2},w_{2} \in \Sigma'_{2}$,
where $\tau$ is defined as
\begin{equation}
\tau \equiv t 
  \left( 2g^{\mathrm{A} (1)}_{p_{1}\bar{p}_{1}} \right)^{\frac{1}{2}}
  \left( 2g^{\mathrm{A} (2)}_{p_{2}\bar{p}_{2}} \right)^{\frac{1}{2}}~.
\label{eq:def-tau}
\end{equation}
Here $G^{\mathrm{A}}_{1}(z_{1};w_{1})$, 
$G^{\mathrm{A}}_{2} (z_{2};w_{2})$
are the Arakelov Green's function with respect to the Arakelov metrics
$g^{\mathrm{A}(1)}_{z_{1} \bar{z}_{1}}$, 
$g^{\mathrm{A}(2)}_{z_{2} \bar{z}_{2}}$
on $\Sigma_{1}$, $\Sigma_{2}$, respectively. 
Taking eq.(\ref{eq:gA-expGA}) into account, we find that
eq.(\ref{eq:G-t}) yields the asymptotic behavior
of the Arakelov metric,
\begin{eqnarray}
2g^{\mathrm{A}}_{z_{1}\bar{z}_{1}} 
 &\sim& |\tau|^{2 \left(\frac{h_{2}}{h} \right)^{2}}
        \, 2g^{\mathrm{A} (1)}_{z_{1}\bar{z}_{1}}
        \, e^{2\frac{h_{2}}{h} G^{\mathrm{A}}_{1} (z_{1},p_{1})}~,
\nonumber \\
2g^{\mathrm{A}}_{z_{2}\bar{z}_{2}}
 &\sim& |\tau|^{2 \left(\frac{h_{1}}{h} \right)^{2}}
        \,2g^{\mathrm{A} (2)}_{z_{2}\bar{z}_{2}}
        \, e^{2\frac{h_{1}}{h} G^{\mathrm{A}}_{2} (z_{2},p_{2})}~.
\label{eq:g-t}
\end{eqnarray}

\subsubsection*{Asymptotics of $Z^{\mathrm{LC}}$}

Let us study the behavior of $Z^{\mathrm{LC}}$,
using the expression (\ref{eq:Z-LC-Arakelov}).
For this purpose, 
we have to know the asymptotic behavior
of the constant $c$
on the degenerating surface $\Sigma$.
This can be obtained by 
substituting eq.(\ref{eq:G-t}) into
the second relation
in eq.(\ref{eq:alpha-Arakelov}) as follows:
\begin{equation}
c (t) \sim  - 2 \frac{h_{1}h_{2}}{h^{2}} \ln |\tau|
         + \frac{h_{1}}{h} c_{1}
         + \frac{h_{2}}{h} c_{2}
\label{eq:c-t}
\end{equation}
as $t \to 0$,
where $c_{1}$ is a constant defined by using 
$g^{\mathrm{A}(1)}_{z\bar{z}}$, $G^{\mathrm{A}}_{1} (z;w)$
and
the Mandelstam mapping $\rho_{1}(z)$
on $\Sigma_{1}$ in the same way as
$c$ defined on $\Sigma$  in eq.(\ref{eq:metric-z-Arakelov}),
and similarly for $c_{2}$ on $\Sigma_{2}$.
Let 
$\alpha_{p_{1}}=-\alpha_{p_{2}}$ be the string-length
of the intermediate propagator in the light-cone string diagram
corresponding to the plumbing fixture:
\begin{equation}
\alpha_{p_{1}} = -\alpha_{p_{2}} = \sum_{r_{2}} \alpha_{r_{2}}
 = -\sum_{r_{1}} \alpha_{r_{1}}~.
\end{equation}
In deriving eq.(\ref{eq:c-t}), we have used
\begin{eqnarray}
\left| \alpha_{p_{1}} \right|^{2}
 &=& \exp \left[ -\sum_{I_{1}} G^{\mathrm{A}}_{1} (z_{I_{1}};p_{1})
                 + \sum_{r_{1}} G^{\mathrm{A}}_{1} (p_{1};Z_{r_{1}})
                + c_{1} \right]~,
\nonumber \\
\left| \alpha_{p_{2}} \right|^{2}
 &=& \exp \left[ -\sum_{I_{2}} G^{\mathrm{A}}_{2} (z_{I_{2}};p_{2})
                 + \sum_{r_{2}} G^{\mathrm{A}}_{2} (p_{2};Z_{r_{2}})
                 + c_{2} \right]~,
\end{eqnarray}
and $|\alpha_{p_{1}} |^{2} = |\alpha_{p_{2}}|^{2}$.

\begin{figure}[h]
\begin{center}
\includegraphics[width=0.65\textwidth]{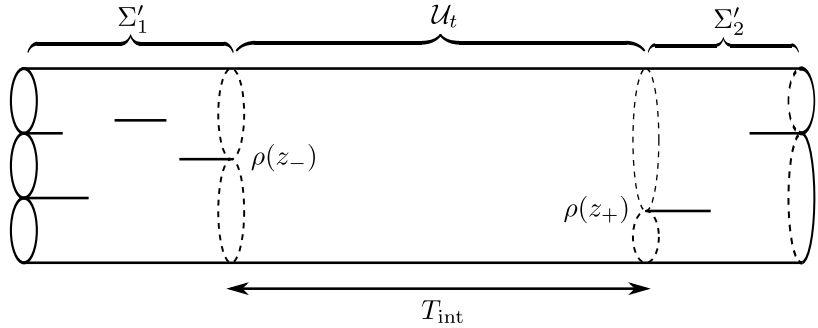}
\end{center}
\caption{The degeneration of the string diagram described by
         the pluming fixture $\mathcal{U}_{t}$. 
         The limit $t\to 0$ corresponds to the limit
         $T_{\mathrm{int}} \to \infty$.}
\label{fig:plumbing2}
\end{figure}

Combined with eq.(\ref{eq:c-t}),
eq.(\ref{eq:metric-z-Arakelov}) yields
\begin{equation}
\partial \rho (z_{1}) \sim \partial \rho_{1} (z_{1})~,
\qquad \partial \rho (z_{2}) \sim \partial \rho_{2} (z_{2})~,
\label{eq:asymptotic-del-rho}
\end{equation}
for $z_{1} \in \Sigma'_{1}$ and $z_{2} \in \Sigma'_{2}$,
as $t \to 0$. 
These can also be obtained from the relations
\begin{eqnarray}
\rho (z_{1})
 &\sim& \rho_{1} (z_{1})
        + \alpha_{m_{1}} \ln (-t)^{-\frac{1}{2}}
        +\sum_{r_{2}} \alpha_{r_{2}} \ln E_{2} (p_{2},Z_{r_{2}})~,
\nonumber \\
\rho (z_{2})
 &\sim& \rho_{2} (z_{2}) 
       + \alpha_{m_{2}} \ln (-t)^{-\frac{1}{2}}
       +\sum_{r_{1}} \alpha_{r_{1}} \ln E_{1} (p_{1},Z_{r_{1}})
 \nonumber \\
 & & \quad {}+ 2\pi i \int^{p_{2}}_{P'_{0}} \omega^{(2)} 
          \frac{1}{\mathop{\mathrm{Im}}\Omega_{2}}
          \left( \sum_{r_{2}} \alpha_{r_{2}} 
                    \mathop{\mathrm{Im}} 
                       \int^{Z_{r_{2}}}_{P'_{0}} \omega^{(2)}
                 + \alpha_{m_{2}} \mathop{\mathrm{Im}}
                        \int^{p_{2}}_{P'_{0}} \omega^{(2)}
          \right)
  \nonumber \\
  & & \quad {}-2\pi i \int^{p_{1}}_{P_{0}} \omega^{(1)}
                \frac{1}{\mathop{\mathrm{Im}} \Omega_{1}}
            \left( \sum_{r_{1}} \alpha_{1}
                    \mathop{\mathrm{Im}} 
                      \int^{Z_{r_{1}}}_{P_{0}} \omega^{(1)}
                  + \alpha_{m_{1}} \mathop{\mathrm{Im}}
                      \int^{p_{1}}_{P_{0}} \omega^{(1)}
            \right)~,
\label{eq:asymptotic-rho}
\end{eqnarray}
up to a purely imaginary constant,
which follow from
the definition~(\ref{eq:Mandelstam}) of 
$\rho (z)$ and the behaviors
of $\omega_{i}$, $E(z,w)$
and $\Omega_{ij}$ on the degenerating surface $\Sigma$.
Eq.(\ref{eq:asymptotic-del-rho}) yields
\begin{equation}
\bar{N}^{r_{1}r_{1}}_{00}
 \sim \bar{N}^{(1) r_{1}r_{1}}_{\hphantom{(1)} \,00} ~,
\qquad
\bar{N}^{r_{2}r_{2}}_{00}
 \sim \bar{N}^{(2) r_{2}r_{2}}_{\hphantom{(2)} \,00} ~,
\label{eq:factorization-Neumann-1}
\end{equation}
as $t \to 0$,
where $\bar{N}^{(1) r_{1}r_{1}}_{\hphantom{(1)} \,00}$ and
$\bar{N}^{(2) r_{2}r_{2}}_{\hphantom{(2)} \,00}$
are zero-modes of the Neumann coefficients 
associated with the punctures $Z_{r_{1}}$ and $Z_{r_{2}}$
on the surfaces $\Sigma_{1}$ and $\Sigma_{2}$.
Let us denote by $\rho (z_{-})$ and $\rho (z_{+})$
the interaction points where the intermediate propagator
corresponding to the plumbing fixture interacts
on $\Sigma'_{1}$ and $\Sigma'_{2}$ respectively,
as is depicted in Figure~\ref{fig:plumbing2}.
Using eq.(\ref{eq:asymptotic-rho}),
we find that
$T_{\mathrm{int}} \equiv 
   \mathop{\mathrm{Re}} \rho ( z_{+} )
   - \mathop{\mathrm{Re}} \rho ( z_{-} )$
is asymptotically related to $t$ by
\begin{equation}
\frac{T_{\mathrm{int}}}{\alpha_{m_{2}}}
\sim
- \ln |t| 
  + \mathop{\mathrm{Re}} 
     \bar{N}^{(1) p_{1} p_{1}}_{\quad \,00}
  + \mathop{\mathrm{Re}} 
     \bar{N}^{(2)p_{2}p_{2}}_{\quad \,00}~,
\end{equation}
where
$\bar{N}^{(1) p_{1} p_{1}}_{ \quad \,00}$,
$\bar{N}^{(2) p_{2} p_{2}}_{ \quad \,00}$
denote the Neumann coefficients associated with
the punctures $p_{1}$, $p_{2}$ on $\Sigma_{1}$, $\Sigma_{2}$,
respectively.

The behavior of the Faltings' invariant
$\delta (\Sigma)$ on the degenerating surface $\Sigma$ 
can be deduced~\cite{Wentworth:1991}
by the use of eqs.(\ref{eq:G-t}) and (\ref{eq:g-t}) as
\begin{equation}
e^{-\frac{1}{4} \delta (\Sigma)} 
\sim |\tau|^{\frac{h_{1}h_{2}}{h}}
     e^{-\frac{1}{4} \delta (\Sigma_{1})}
     e^{-\frac{1}{4} \delta (\Sigma_{2})}~.
\label{eq:Faltings-t}
\end{equation}

Gathering all the results obtained above,
we eventually find that on the degenerating surface
$Z^{\mathrm{LC}}$
factorizes as
\begin{equation}
Z^{\mathrm{LC}}
 \sim \frac{\mathcal{C}_{h,N}(\beta_A)}
           {\mathcal{C}_{h_{1},N_{1}} (\beta_{A_1})
            \mathcal{C}_{h_{2},N_{2}}(\beta_{A_2})}
      \, e^{\frac{2T_{\mathrm{int}}}{\alpha_{m_{2}}}}
      \, Z^{\mathrm{LC}}_{1}
         Z^{\mathrm{LC}}_{2}~,
\label{eq:ZX-expGamma-t}
\end{equation}
where $Z^{\mathrm{LC}}_{1}$ and $Z^{\mathrm{LC}}_{2}$
are the partition functions on $\Sigma_{1}$
and $\Sigma_{2}$ respectively.
In order that 
$Z^{\mathrm{LC}}$ should correctly factorize as 
\begin{equation}
Z^{\mathrm{LC}}
 \sim 
      \, e^{\frac{2T_{\mathrm{int}}}{\alpha_{m_{2}}}}
      \, Z^{\mathrm{LC}}_{1}
         Z^{\mathrm{LC}}_{2}~,
\label{eq:correct}
\end{equation}
$\mathcal{C}_{h,N}(\beta_{A} )$
has to satisfy
\begin{equation}
\frac{\mathcal{C}_{h,N}(\beta_{A} )}
     {\mathcal{C}_{h_{1},N_{1}}(\beta_{A_1} ) 
       \mathcal{C}_{h_{2},N_{2}}(\beta_{A_2})}
=1~.
\label{eq:cond-C-1}
\end{equation}
Repeating the same procedure, we can see that 
the evaluation of $\mathcal{C}_{h,N}$ 
reduces to that of $\mathcal{C}_{1,N}$.

\subsubsection{$h=1$ case}

For $h=1$, it is convenient to define a complex coordinate $u$ 
on $\Sigma$ such that
\begin{equation}
du = \omega~,
\end{equation}
where $\omega$ is the unique holomorphic one-form satisfying
\begin{equation}
\oint_a\omega =1\,,
\qquad \oint_b\omega =\tau \,.
\end{equation}

In terms of the coordinate $u$ and the moduli parameter $\tau$, 
the prime form $E(u,u^\prime )$ takes the form
\begin{equation}
E(u,u^\prime ) = \frac{\theta_{1} (u-u^\prime |\tau)}{\theta'_{1} (0|\tau)}~.
\end{equation}
Here $\theta_{1} (u|\tau)$ denotes the theta function for $h=1$
with the odd spin structure, defined as
\begin{equation}
\theta_{1} (u|\tau)
 \equiv - \theta {1/2 \atopwithdelims[] 1/2} (u|\tau)
 = -e^{\frac{i}{4} \pi \tau + i\pi \left(u+\frac{1}{2}\right)}
    \theta \left. \left(u+\frac{\tau}{2}+\frac{1}{2} \right|\tau \right)~,
\label{eq:1-loop-theta}
\end{equation}
which is related to the Dedekind eta function $\eta (\tau)$ by
$\theta'_{1} (0|\tau) = 2\pi \eta(\tau)^{3}$.
Accordingly, the Mandelstam mapping $\rho (u)$ becomes
\begin{equation}
\rho (u) = \sum_{r=1}^{N} \alpha_{r}
   \left[ \ln \theta_{1} (u-U_{r}|\tau)
          -2\pi i \frac{\mathop{\mathrm{Im}} U_{r}}
                       {\mathop{\mathrm{Im}} \tau}
                (u-u_{0})
   \right],
\label{eq:Mandelstam-torus}
\end{equation}
where $U_{r}$ $(r=1,\ldots, N)$ denote the punctures 
and $u_{0}$ denotes the base point on the $u$-plane.
Let $u_{I}$ $(I=1,\ldots,N)$ be the interaction points on the $u$-plane, 
determined by
$\partial \rho (u_{I})=0$.
For the $h=1$ Riemann surface~$\Sigma$
that we are considering,
eq.(\ref{eq:divisor-rho}) tells us that there exist
integers $m$ and $n$ such that
\begin{equation}
\sum_{I=1}^{N} u_{I} - \sum_{r=1}^{N} U_{r}
 = m + n \tau~.
\label{eq:Abel-thm}
\end{equation}
The Arakelov metric $g_{u\bar{u}}^\mathrm{A}$ 
does not depend on $u,\bar{u}$ 
because of eq.(\ref{eq:ArakelovR}) and the Arakelov Green's function 
$G^\mathrm{A}(u;u^\prime )$ is given as 
\begin{equation}
G^\mathrm{A}(u;u^\prime )
=
-\ln\left|
\frac{\theta_{1} (u-u^\prime |\tau)}{\theta'_{1} (0|\tau)}
\right|^2
+
\frac{2\pi}{\mathop{\mathrm{Im}} \tau}
\left( \mathop{\mathrm{Im}} (u-u^\prime )\right)^2
-\ln (2g_{u\bar{u}}^\mathrm{A})\,.
\end{equation} 
Substituting all these into eq.(\ref{eq:Z-LC-Arakelov}), we obtain
\begin{equation}
Z^{\mathrm{LC}}
 = \mathcal{C}_{1,N}(\beta ) (2\pi )^{-16}
    ( \mathop{\mathrm{Im}} \tau )^{-12}
   \left| \eta (\tau )\right|^{-48}
   \prod_{r=1}^{N} \left( e^{-2\mathop{\mathrm{Re}} \bar{N}^{rr}_{00}}
                    \alpha_{r}^{-2} \right)
   \prod_{I=1}^{N} \left| \partial^{2} \rho (u_{I}) \right|^{-1} \,.
\end{equation}

\begin{figure}[htbp]
\begin{center}
\includegraphics[width=0.875\textwidth]{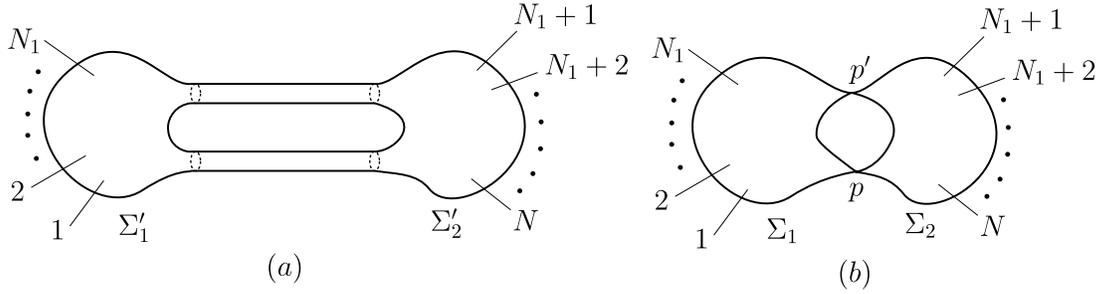}
\end{center}
\caption{The degeneration of the $h=1$ Riemann surface}
\label{fig:h=1}
\end{figure}

$\mathcal{C}_{1,N}(\beta )$ can be fixed 
by considering the degeneration of the
Riemann surface~$\Sigma$ of the type depicted in Figure~\ref{fig:h=1}.
In this degeneration, two non-trivial cycles,
the $a$ cycle and its complement, are pinched to
nodes $p$ and $p'$ respectively, and
the genus $1$ surface $\Sigma$ will be divided into two disconnected
spheres $\Sigma_{1}$ and $\Sigma_{2}$.
We denote the nodes $p,p'$ by
$p_{1},p'_{1}$ or $p_{2},p'_{2}$
depending on whether they are regarded
as punctures in $\Sigma_{1}$ or those in $\Sigma_{2}$.
Similarly to the $h>1$ case, 
we assume that
the set of the punctures $U_{r}$ $(r=1,\ldots,N)$
are ordered so that in this degeneration
they will be divided into
two groups as $U_{r} = (U_{r_{1}},U_{r_{2}})$
$(1\leq r_{1} \leq N_{1},\, N_{1}+1 \leq r_{2} \leq N)$
with $U_{r_{1}} \in \Sigma'_{1}$ and $U_{r_{2}} \in \Sigma'_{2}$.
We may similarly assume that 
the set of the interaction points $u_{I}$ $(I=1,\ldots,N)$
will be divided as $u_{I}=(u_{I_{1}},u_{I_{2}})$
$(1\leq I_{1} \leq N_{1},\,N_{1}+1 \leq I_{2} \leq N)$
with $u_{I_{1}} \in \Sigma'_{1}$ and $u_{I_{2}} \in \Sigma'_{2}$.
%
%
%
%
The resultant spheres $\Sigma_{1}$ and $\Sigma_{2}$ are 
with $N_{1}+2$ punctures $(U_{r_{1}},p_{1},p'_{1})$
and $N_{2}+2$ punctures $(U_{r_{2}},p_{2},p'_{2})$,
respectively.

\begin{figure}[htbp]
\begin{center}
\includegraphics[width=0.6\textwidth]{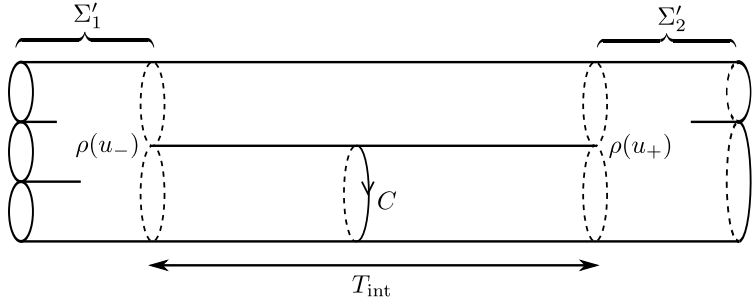}
\end{center}
\caption{The degeneration of the light-cone string diagram.
         This corresponds to the limit $T_{\mathrm{int}} \to \infty$}
\label{fig:h=1plumbing}
\end{figure}

The degeneration mentioned above is achieved by
taking the limit in which the heights of the two cylinders
corresponding to the internal propagators
composing the loop are infinitely long (Figure~\ref{fig:h=1}~(a)).
In the light-cone string diagram, this corresponds to the limit
$T_\mathrm{int}\to\infty$ as depicted in Figure~\ref{fig:h=1plumbing},
which is the limit $\mathop{\mathrm{Im}}\tau\to\infty$
with the lengths
of the two intermediate strings fixed. 
The length of one of the strings is proportional to 
\begin{eqnarray}
\oint_{C}du \, \partial\rho\left(u\right)
&=&
\rho\left(u+1\right)-\rho\left(u\right)
\nonumber
\\
&=&
-2\pi i
\frac{\sum_{r} \alpha_{r} \mathop{\mathrm{Im}}U_{r}}
     {\mathop{\mathrm{Im}} \tau}
+\sum_{r}\left(\pm\pi i\right)\alpha_{r}\,.
\end{eqnarray}
The limit we take is
\begin{eqnarray}
U_{r_{1}},u_{I_{1}} 
 &\sim & 
   \mathcal{O} \left( \left(\mathop{\mathrm{Im}} \tau
                      \right)^{0}
               \right)~,
\nonumber \\
U_{r_{2}} 
 & = &  R\tau+U_{r_{2}}^{\prime}~,
\qquad \mbox{with} \quad
  U_{r_{2}}^{\prime} 
   \sim  \mathcal{O}\left( \left( \mathop{\mathrm{Im}} \tau
                   \right)^{0} \right)~,
\nonumber \\
u_{I_{2}} 
  & = & R\tau + u_{I_{2}}^{\prime}~,
\qquad \, \mbox{with} \quad
  u_{I_{2}}^{\prime} 
   \sim  \mathcal{O}\left( \left( \mathop{\mathrm{Im}} \tau
                    \right)^{0} \right)~,
\label{eq:u-limit}
\end{eqnarray}
where $R$ is a real parameter such that $0<R<1$.
We keep 
\begin{equation}
\sum_{r_{1}}\alpha_{r_{1}} \mathop{\mathrm{Im}} U_{r_{1}}
 +\sum_{r_{2}}
     \alpha_{r_{2}} \mathop{\mathrm{Im}} U_{r_{2}}^{\prime} 
 =0 
\label{eq:ImUr}
\end{equation}
in taking the limit $\mathop{\mathrm{Im}}\tau\to\infty$ 
to make
\begin{equation}
\frac{\sum_{r}\alpha_{r} \mathop{\mathrm{Im}} U_{r}}
     {\mathop{\mathrm{Im}} \tau}
  =  R\sum_{r_{2}}\alpha_{r_{2}}
\end{equation}
fixed.
Substituting eq.(\ref{eq:u-limit}) into eq.(\ref{eq:Abel-thm})
yields $n=0$ and thus
\begin{equation}
\sum_{I=1}^{N} u_{I} - \sum_{r=1}^{N} U_{r}
 = m~,
\qquad m \in \mathbb{Z}~.
\label{eq:Abel-thm2}
\end{equation}

It is straightforward to show that
in the degeneration limit $\mathop{\mathrm{Im}}\tau \to \infty$
addressed above,
for 
$u\sim\mathcal{O}\left(\left( \mathop{\mathrm{Im}} \tau
                       \right)^{0}\right)$,
defining
\begin{equation}
z\equiv e^{-2\pi iu}~,  
\qquad
Z_{r_{1}} \equiv  e^{-2\pi iU_{r_{1}}}~,
\qquad
Z'_{r_{2}} \equiv e^{-2\pi i U'_{r_{2}}}~,
\end{equation}
we have
\begin{eqnarray}
\rho (u) 
 &\sim &  \rho_{1}(z)
   -\pi i R  \tau  \sum_{r_{2}} \alpha_{r_{2}} 
  -\frac{1}{2} \sum_{r_{1}} \alpha_{r_{1}} \ln Z_{r_{1}}
  + \frac{1}{2} \sum_{r_{2}} \alpha_{r_{2}} \ln Z'_{r_{2}}
\nonumber \\
& & 
\ \qquad
 {}+\sum_{r_{2}} \left( \pm \pi i \right) \alpha_{r_{2}}
   + 2 \pi  i R  u_{0}  \sum_{r_{2}} \alpha_{r_{2}}~.
\label{eq:rho-limit-1}
\end{eqnarray}
Here $\rho_{1} (z)$ is defined as
\begin{equation}
\rho_{1} (z)
 \equiv  
   \sum_{r_{1} =1}^{N_{1}}
       \alpha_{r_{1}} \ln \left(z-Z_{r_{1}}\right)
   + \left( R \sum_{r_{2}}\alpha_{r_{2}} \right) \ln z~,
\end{equation}
which coincides with the Mandelstam
mapping on the sphere $\Sigma_{1}$
with parameters
$Z_{p_{1}}=0$,
$\alpha_{p_{1}}
  =R \sum_{r_{2}}\alpha_{r_{2}}$ for puncture $p_{1}$
and 
$Z_{p'_{1}}=\infty$,
$\alpha_{p'_{1}}
  =(R-1) \sum_{r_{1}} \alpha_{r_{1}}$
for puncture $p'_{1}$.

On the other hand, for $u=R\tau+u^{\prime}$ with 
$u^{\prime} 
  \sim \mathcal{O} \left(\left(\mathop{\mathrm{Im}}\tau
                   \right)^{0}\right)$,
introducing
\begin{equation}
z^{\prime} \equiv  e^{-2\pi iu^{\prime}}~,
\end{equation}
we obtain
\begin{eqnarray}
\rho (u) 
 &\sim& \rho_{2} (z')
     -2\pi i \left( R^{2} - \frac{R}{2} \right) \tau
             \sum_{r_{2}} \alpha_{r_{2}}
\nonumber \\
 & & 
\ \qquad
{}- \frac{1}{2} \sum_{r_{1}} \alpha_{r_{1}} \ln Z_{r_{1}}
     - \frac{1}{2} \sum_{r_{2}} \alpha_{r_{2}} \ln Z'_{r_{2}}
  + 2\pi i R u_{0}
         \sum_{r_{2}} \alpha_{r_{2}}~,
\label{eq:rho-limit-2}
\end{eqnarray}
where $\rho_{2} (z')$ is defined as
\begin{equation}
\rho_{2} \left( z' \right)
  \equiv 
    \sum_{r_{2}=1}^{N_{2}}
       \alpha_{r_{2}}
       \ln \left(z^{\prime}-Z_{r_{2}}^{\prime}\right)
    + \left( (R-1) \sum_{r_{2}} \alpha_{r_{2}} \right)
        \ln z^{\prime}~,
\end{equation}
which coincides with the Mandelstam
mapping on the sphere $\Sigma_{2}$
with parameters
$Z'_{p_{2}}=\infty$,
$\alpha_{p_{2}}
  \left( = - \alpha_{p_{1}} \right)
  =R\sum_{r_{1}}\alpha_{r_{1}}$
for puncture $p_{2}$
and 
$Z'_{p'_{2}}=0$,
$\alpha_{p'_{2}}
  \left( = - \alpha_{p'_{1}} \right)
  = (R-1) \sum_{r_{2}}\alpha_{r_{2}}$
for puncture $p'_{2}$.

It follows from eqs.(\ref{eq:rho-limit-1}) 
and (\ref{eq:rho-limit-2})
that
in the degeneration limit,
the Neumann coefficients
$\bar{N}^{r_{1} r_{1}}_{00}$, $\bar{N}^{r_{2}r_{2}}_{00}$
behave as
\begin{equation}
\bar{N}^{r_{1}r_{1}}_{00} 
  \sim  \bar{N}^{(1) r_{1}r_{1}}_{\quad 00}
          -\ln (-2\pi i)
          + 2\pi i U_{r_{1}}~,
\qquad
\bar{N}^{r_{2}r_{2}}_{00} 
 \sim \bar{N}^{(2) r_{2}r_{2}}_{\quad 00}
       - \ln (-2\pi i)
          + 2\pi i U'_{r_{2}}~,
\end{equation}
where $\bar{N}^{(1)r_{1}r_{1}}_{\quad 00}$,
$\bar{N}^{(2)r_{2}r_{2}}_{\quad 00}$
denote the Neumann coefficients
on $\Sigma_{1}$, $\Sigma_{2}$
associated with the punctures
$Z_{r_{1}}$, $Z'_{r_{2}}$ respectively,
and that $\partial^{2} \rho (u_{I_{1}})$,
$\partial^{2} \rho (u_{I_{2}})$ behave as
\begin{equation}
\partial^{2} \rho (u_{I_{1}})
\sim -(2\pi)^{2} e^{-4\pi i U_{I_{1}}} 
     \partial^{2} \rho_{1} (z_{I_{1}})~,
\qquad
\partial^{2} \rho (u_{I_{2}}) 
\sim  -(2\pi)^{2} e^{-4\pi i U'_{I_{2}}} 
     \partial^{2} \rho_{2} (z'_{I_{2}})~,
\end{equation}
where
\begin{equation}
z_{I_{1}} \equiv e^{-2\pi i u_{I_{1}}}~,
\qquad
z'_{I_{2}} \equiv e^{-2\pi i u'_{I_{2}}}~.
\end{equation}

Let $\rho (u_{-})$ and $\rho (u_{+})$ be
the interaction points in $\Sigma'_{1}$ and $\Sigma'_{2}$
on the light-cone string diagram
where the long intermediate propagators
interact as described in Figure~\ref{fig:h=1plumbing}.
Using eq.(\ref{eq:ImUr}),
we can derive from
eqs.(\ref{eq:rho-limit-1}) and (\ref{eq:rho-limit-2})
that 
$T_{\mathrm{int}} 
  \equiv \mathop{\mathrm{Re}} \rho (u_{+}) 
         -  \mathop{\mathrm{Re}} \rho (u_{-})$
is asymptotically related to $\mathop{\mathrm{Im}} \tau$
as
\begin{equation}
\left( \frac{1}{\alpha_{p_{2}}} 
       + \frac{1}{\alpha_{p'_{2}}} \right) T_{\mathrm{int}}
\sim  2\pi \mathop{\mathrm{Im}} \tau
        + \mathop{\mathrm{Re}} 
          \left( \bar{N}^{(1)p_{1}p_{1}}_{\quad 00}
                  + \bar{N}^{(1) p'_{1}p'_{1}}_{\quad 00}
                  + \bar{N}^{(2) p_{2}p_{2}}_{\quad 00}
                  + \bar{N}^{(2) p'_{2}p'_{2}}_{\quad 00}
          \right)~,
\end{equation}
where $\bar{N}^{(1)p_{1}p_{1}}_{\quad 00}$,
$\bar{N}^{(1)p'_{1}p'_{1}}_{\quad 00}$
are the Neumann coefficients on $\Sigma_{1}$
associated with the punctures
$p_{1}$, $p'_{1}$,
and similarly for $\bar{N}^{(2)p_{2}p_{2}}_{\quad 00}$,
$\bar{N}^{(2) p'_{2}p'_{2}}_{\quad 00}$
with the roles of $\Sigma_{1}$ and $\Sigma_{2}$
interchanged.

Gathering all the results obtained above
and using the behavior
$
\left| \eta (\tau) \right|^{-48}
 \sim e^{4\pi \mathop{\mathrm{Im}} \tau}
$
as $\mathop{\mathrm{Im}} \tau \to \infty$,
we conclude that on the degenerating surface,
the partition function $Z^{\mathrm{LC}}$
behaves as
\begin{equation}
Z^\mathrm{LC}
\sim
\mathcal{C}_{1,N}(\beta ) \left( 32\pi^2 \right)^4
\left(8\pi^2 \mathop{\mathrm{Im}} \tau \right)^{-12}
\exp\left( \frac{2T_{\mathrm{int}}}{\alpha_{p_{2}}}
            + \frac{2T_{\mathrm{int}}}{\alpha_{p'_{2}}}\right)
Z^{\mathrm{LC}}_{1}  Z^{\mathrm{LC}}_{2}~,
\label{eq:ZLCh=1}
\end{equation}
where $Z^{\mathrm{LC}}_{1}$, $Z^{\mathrm{LC}}_{2}$ are the tree-level 
partition functions given in Ref.~\cite{Baba:2009ns}
\begin{equation}
Z^{\mathrm{LC}}
=
\left| \sum_{s=1}^{N} \alpha_{s} Z_{s} \right|^{4}
\prod_{r} \left( e^{-2\mathop{\mathrm{Re}} \bar{N}^{rr}_{00}}
                 \alpha_{r}^{-2}
          \right)
\prod_{I} \left| \partial^{2} \rho (z_{I}) \right|^{-1}
\end{equation}
for the surfaces $\Sigma_1, \Sigma_2$
with $Z_{p'_{1}} \to \infty$ for $Z^{\mathrm{LC}}_{1}$
and $Z'_{p_{2}} \to \infty$ for $Z^{\mathrm{LC}}_{2}$.
The factor 
 $\left(8\pi^2 \mathop{\mathrm{Im}}\tau \right)^{-12}$ 
on the right hand side of eq.(\ref{eq:ZLCh=1}) 
coincides with that from the integration over the loop momentum and 
$\exp\left( \frac{2T_{\mathrm{int}}}{\alpha_{p_{2}}}
            +\frac{2T_{\mathrm{int}}}{\alpha_{p'_{2}}}
     \right)$ 
can be identified with the contribution from the tachyon mass. 
Therefore by taking 
\begin{equation}
\mathcal{C}_{h=1,N}(\beta ) 
 = \frac{1}{(32\pi^{2})^{4}}~,
\label{eq:initial-cond}
\end{equation}
we get the factorization property as desired. 
Thus we find that $\mathcal{C}_{h=1,N}(\beta )$ is 
just a numerical constant. 
With this equation taken as initial condition,
we can inductively solve eq.(\ref{eq:cond-C-1}) for 
$\mathcal{C}_{h,N}(\beta_A)$
and obtain
\begin{equation}
\mathcal{C}_{h,N}(\beta_A )
 = \frac{1}{(32\pi^{2})^{4h}}~.
\label{eq:result-C}
\end{equation}
We eventually get $Z^\mathrm{LC}$ as is given
in eq.(\ref{eq:ZLCexpression}).


\section{Modular transformations}
\label{sec:modular-prop}

In this appendix, we would like to 
show that
the partition function $Z^{\mathrm{LC}}$ and 
the correlation function
$\left\langle \prod_{r=1} V^{\mathrm{LC}}_{r} \right\rangle^{X^{i}}$
are modular invariant respectively.

We first review the modular properties of
mathematical quantities on the surface $\Sigma$
to fix the notations.
Suppose that $M \in Sp(2h,\mathbb{Z})$,
namely $M$ is a $2h\times 2h$ integral matrix satisfying
\begin{equation}
M J M^{T} =J~,
\qquad J \equiv 
 \left( \begin{array}{cc} 0 & 1_{h} \\ -1_{h} &0 \end{array} \right)~, 
\end{equation}
where $1_{h}$ denotes the $h\times h$ unit matrix.
Decomposing the matrix $M$
as
$M=\left( \begin{array}{cc} A & B \\ C &D \end{array} \right)$
with $A$, $B$, $C$, $D$ being the $h\times h$ matrices,
one can show that these matrices satisfy
\begin{equation}
\left\{ \begin{array}{l}
         AD^{T} - BC^{T} =1_{h}\\
         AB^{T} = BA^{T}\\
         CD^{T} = DC^{T}
        \end{array}
\right.~,\quad
\left\{ \begin{array}{l}
          A^{T}D-C^{T}B=1_{h}\\
          A^{T}C=C^{T}A\\
          B^{T}D=D^{T}B
        \end{array}
\right.~,\quad
\left( \begin{array}{cc}
          D & C \\ B&A
       \end{array}
\right) \in Sp(2h,\mathbb{Z})~.
\label{eq:Sp2h-relations}
\end{equation}

Let us consider the modular transformation
under which the homology basis $\{a_{i},b_{i}\}$
($i=1,\ldots,h)$
transforms as
\begin{equation}
\left( \begin{array}{c} a \\ b \end{array} \right)
\mapsto
\left( \begin{array}{c} \tilde{a} \\ \tilde{b}\end{array}
\right) 
=\left( \begin{array}{cc} D&C \\ B &A \end{array} \right)
\left( \begin{array}{c} a \\ b \end{array} \right)~.
\label{eq:modular1}
\end{equation}
Under this transformation,
$\omega$ and $\Omega$ respectively transform as
\begin{equation}
\omega \mapsto 
 \tilde{\omega} = \omega \frac{1}{C\Omega +D}~,
\qquad
\Omega \mapsto
 \tilde{\Omega} = (A\Omega +B)\frac{1}{C\Omega +D}~.
\label{eq:modular-omega}
\end{equation}
For the theta function with spin structure
$[s] = {s' \atopwithdelims[] s''}$,
there is the following transformation law~\cite{Igusa:1972}:
\begin{equation}
\theta [\tilde{s}] \left. \!
  \left( \tilde{\zeta} \right| \tilde{\Omega} \right)
 = \varepsilon (M) e^{i\pi\phi (s)}
   \det \left( C\Omega +D \right)^{\frac{1}{2}}
   \exp\left( i\pi \zeta \frac{1}{C\Omega +D} C\zeta \right)
   \theta [s] (\zeta|\Omega)~,
\label{eq:modular-theta}
\end{equation}
where $\varepsilon (M)$ is an eighth root of unity
depending on $M$,
\begin{eqnarray}
&&
\tilde{s} \equiv \tilde{s}_{0} +\delta~,
\qquad
 [\tilde{s}_{0}] 
    = { \tilde{s}'_{0} \atopwithdelims[] \tilde{s}''_{0}}~,
\qquad
 [\delta]
  = {\delta' \atopwithdelims[] \delta''}~,
\nonumber \\[1ex]
&&
\left( \begin{array}{c} 
          \tilde{s}'_{0} \\ \tilde{s}''_{0}
       \end{array} 
\right)
 = \left( \begin{array}{cc} D & -C \\ -B & A \end{array} \right)
   \left( \begin{array}{c} s' \\ s'' \end{array} \right)~,
\quad
 \left( \begin{array}{c} \delta' \\ \delta'' \end{array}\right)
      = \frac{1}{2} \left( 
          \begin{array}{c}
             \mathrm{diag} (CD^{T}) \\ \mathrm{diag} (AB^{T})
          \end{array}
          \right)~,
\label{eq:modular-characteristics}
\end{eqnarray}
and
\begin{equation}
\tilde{\zeta} = \zeta \frac{1}{C\Omega +D}~,
\qquad 
\phi (s) = \tilde{s}'_{0} \tilde{s}''_{0}
            - s's'' + 2\tilde{s}'_{0} \delta''~.
\label{eq:modular-zeta}
\end{equation}
The transformation law~(\ref{eq:modular-theta})
leads to 
\begin{equation}
\Delta \mapsto \tilde{\Delta}
  \equiv \Delta \frac{1}{C\Omega +D} 
    + \delta' \tilde{\Omega} + \delta''
\qquad \pmod{\mathbb{Z}^{h} + \mathbb{Z}^{h}\Omega}~.
\label{eq:modular-delta}
\end{equation}
This is an immediate result of the relation,
\begin{eqnarray}
\theta \left. \! \left( \tilde{\zeta} + \delta' \tilde{\Omega}
                        + \delta''
       \right| \tilde{\Omega} \right)
 &=& \exp \left[ -i\pi \delta' \tilde{\Omega} \delta'
                 -i 2\pi \zeta \frac{1}{C\Omega +D} \delta'
                 + i\pi \zeta \frac{1}{C\Omega +D} C \zeta
          \right]
 \nonumber \\
 & & \times \varepsilon (M) e^{-i2\pi \delta' \delta''}
     \det \left( C\Omega +D \right)^{\frac{1}{2}}
     \theta (\zeta | \Omega )~,
\label{eq:tilde-theta-theta}
\end{eqnarray}
which is obtained from eq.(\ref{eq:modular-theta})
by setting $s=0$ and using eq.(\ref{eq:theta-char}).

Using eq.(\ref{eq:Sp2h-relations}), we can show
that the matrices $\tilde{\Omega}$
and $\frac{1}{C\Omega +D} C$ are symmetric,
\begin{eqnarray}
\mathop{\mathrm{Im}} \tilde{\Omega}
 = \frac{1}{\bar{\Omega} C^{T} +D^{T}} 
   \mathop{\mathrm{Im}} \Omega
   \frac{1}{C\Omega +D}~,
\label{eq:Sp2-relations2}
\end{eqnarray}
and thereby
\begin{equation}
\mathop{\mathrm{Im}} \left( v \frac{1}{C\Omega +D} \right)
  \frac{1}{\mathop{\mathrm{Im}} \tilde{\Omega}}
  \mathop{\mathrm{Im}} \left( \frac{1}{\Omega C^{T}+D^{T}} v
                       \right)
 = \mathop{\mathrm{Im}} v \frac{1}{\mathop{\mathrm{Im}} \Omega}
   \mathop{\mathrm{Im}} v
   - \mathop{\mathrm{Im}}
      \left( v \frac{1}{C\Omega +D} Cv \right)
\label{eq:Sp2-relations3}
\end{equation}
for an arbitrary vector $v \in \mathbb{C}^{h}$.
This relation is useful in the following calculation.

Now let us show 
the modular invariance of 
$Z^{\mathrm{LC}}
  =Z^{X}[g^{\mathrm{A}}_{z\bar{z}}]^{24}
    e^{-\Gamma \left[g^{\mathrm{A}}_{z\bar{z}},
      \, \ln |\partial \rho|^{2}\right]}$.
We will show that each of $Z^{X}[g^{\mathrm{A}}_{z\bar{z}}]$
and $e^{-\Gamma \left[g^{\mathrm{A}}_{z\bar{z}},
        \, \ln |\partial \rho|^{2}\right]}$
is modular invariant by itself.
First we study the modular transformations of 
$e^{-\Gamma \left[ g^{\mathrm{A}}_{z\bar{z}},\, \ln |\partial \rho|^{2}
            \right]}$.
Using eq.(\ref{eq:modular-theta}), one can find that
the prime form 
$E(z,w)$ transforms as
\begin{equation}
E(z,w) 
 \mapsto 
 \tilde{E} (z,w) 
  = \exp \left[i\pi \int^{z}_{w} \omega \frac{1}{C\Omega +D} C
                    \int^{z}_{w} \omega \right] E(z,w)~.
\label{eq:modular-E-sigma}
\end{equation}
This yields that $F(z,\bar{z};w,\bar{w})$ 
defined in eq.(\ref{eq:propagator2}) is modular
invariant,
and the Mandelstam mapping $\rho (z)$ given in eq.(\ref{eq:Mandelstam})
just shifts by a factor independent of $z$ as
\begin{equation}
\rho (z) \mapsto \tilde{\rho} (z)
 =\rho (z)
  + i\pi \sum_{r=1}^{N} \alpha_{r} \int^{Z_{r}}_{P_{0}} \omega
    \frac{1}{C\Omega +D} C \int^{Z_{r}}_{P_{0}} \omega~.
\label{eq:modular-rho}
\end{equation}
This leads to the modular invariance
of $\partial \rho (z)$ and $\bar{N}^{rr}_{00}$.\footnote{We here 
assume that the local coordinate $z$ is defined in a modular invariant way.}
Taking into account the modular invariance
of $F(z,\bar{z};w,\bar{w})$ and 
$\mu_{z\bar{z}}$ defined in eq.(\ref{eq:Bergman}),
one can show by the use of eqs.(\ref{eq:ArakelovR}) and (\ref{eq:G-B})
that
$g^{\mathrm{A}}_{z\bar{z}}$ and
$G^{\mathrm{A}} (z;w)$
are modular invariant.
Now that we have found that
$g^{\mathrm{A}}_{z\bar{z}}$, $G^{\mathrm{A}} (z;w)$,
$\partial \rho (z)$ and $\bar{N}^{rr}_{00}$
are modular invariant,
it is evident that 
$e^{-\Gamma \left[g^{\mathrm{A}}_{z\bar{z}},
                  \, \ln |\partial \rho|^{2}\right]}$
given in eq.(\ref{eq:Gamma}) 
is modular invariant as well.

Next, let us show that $Z^{X}[g^{\mathrm{A}}_{z\bar{z}}]$
is modular invariant.
This is a direct result of the modular invariance of
the Faltings' invariant $\delta (\Sigma)$
given in eq.(\ref{eq:Faltings}),
which can be seen as follows.
For an arbitrary $\zeta \in \mathbb{C}^{h}$,
the following relation holds:
\begin{eqnarray}
\lefteqn{
e^{-\pi \mathop{\mathrm{Im}} \left( \tilde{\zeta}+ \tilde{\Delta} \right)
        \frac{1}{\mathop{\mathrm{Im}} \tilde{\Omega}}
         \mathop{\mathrm{Im}} \left( \tilde{\zeta} + \tilde{\Delta} \right)}
 \left| \theta \left. \! \left( \tilde{\zeta} + \tilde{\Delta} \right|
                   \tilde{\Omega} \right)
 \right|
} \nonumber \\
&& = \left| \det \left( C \Omega+D\right) \right|^{\frac{1}{2}}
 e^{-\pi \mathop{\mathrm{Im}} \left( \zeta +  \Delta \right)
        \frac{1}{\mathop{\mathrm{Im}} \Omega}
         \mathop{\mathrm{Im}} \left( \zeta + \Delta \right)}
 \left| \theta \left. \! \left( \zeta + \Delta \right|
                   \Omega \right)
 \right|~,
\label{eq:modular-theta3}
\end{eqnarray}
where 
$\tilde{\zeta}$ and $\tilde{\Delta}$
are defined in eqs.(\ref{eq:modular-zeta})
and (\ref{eq:modular-delta}) respectively.
This can be obtained
by using eq.(\ref{eq:tilde-theta-theta})
with $\zeta$ replaced by $\zeta + \Delta$ and
the relation
\begin{equation}
\tilde{\zeta} + \tilde{\Delta}
  \equiv \left( \zeta + \Delta \right) \frac{1}{C\Omega +D}
    + \delta' \tilde{\Omega} + \delta''
\qquad \pmod{\mathbb{Z}^{h} + \mathbb{Z}^{h} \Omega}~.
\label{eq:tilde-zeta-delta}
\end{equation}
Using eqs.(\ref{eq:modular-theta3}),
(\ref{eq:modular-omega}), (\ref{eq:Sp2-relations2}), 
(\ref{eq:abs-theta-char})
and the modular invariance of 
$g^{\mathrm{A}}_{z\bar{z}}$ and $G^{\mathrm{A}} (z;w)$,
one can find that 
$\delta (\Sigma)$ is modular invariant.


Finally we consider the modular invariance of the correlation
function
$\left\langle 
    \prod_{r=1}^{N} V^{\mathrm{LC}}_{r} 
 \right\rangle^{X^{i}}$.
{}From the modular transformation law~(\ref{eq:modular-rho})
of 
$\rho (z)$,
one finds that the local coordinates $w_{r}$ 
defined in eq.(\ref{eq:localcoord}) is modular invariant.
Eq.(\ref{eq:modular-rho}) also leads to
the modular invariance of $\prod_{r}e^{-p^{-}_{r} \tau_{0}^{(r)}}$
contained in $ \prod_{r} V_{r}^{\mathrm{LC}}$,
in the presence of the delta-function
$\delta \left( \sum_{r} p^{-}_{r} \right)$
responsible for the conservation of the momentum $p^{-}$.
These imply that $\prod_{r} V^{\mathrm{LC}}_{r}$
in the correlation function is modular invariant.
We also note that the scalar Green's function
in the worldsheet theory can be described
by using the worldsheet metric and 
modular invariant $F(z,\bar{z};w,\bar{w})$~\cite{Verlinde:1986kw}.
Since we choose the modular invariant $g^{\mathrm{A}}_{z\bar{z}}$
as the worldsheet metric, the scalar Green's function
is modular invariant as well.
Thus we conclude that the correlation function
$\left\langle 
    \prod_{r=1}^{N} V^{\mathrm{LC}}_{r} 
 \right\rangle^{X^{i}}$
is modular invariant.

Putting the results obtained above together,
we find that the amplitudes~(\ref{eq:amplitude})
of the light-cone gauge string field theory in noncritical
dimensions
are modular invariant.


\section{A derivation of eq.(\ref{eq:ghostZX})}
\label{sec:proof-ZLC-Zgh}

In this appendix, we derive the identity~(\ref{eq:ghostZX}):
\begin{eqnarray}
&&
    \prod_{r=1}^{N}
   \left( \alpha_{r} e^{2 \mathop{\mathrm{Re}} \bar{N}^{rr}_{00}} \right)
    e^{-\Gamma  \left[g_{z\bar{z}}^\mathrm{A},\, \ln |\partial \rho|^{2}
              \right]}
   Z^{X} [g_{z\bar{z}}^\mathrm{A}]^{-2}
\nonumber \\
&& = \mbox{const.}
\int\left[dbd\tilde{b}dcd\tilde{c}\right]_{g_{z\bar{z}}^\mathrm{A}}
e^{-S^{bc}}
\prod_{r=1}^{N}c\tilde{c}(Z_r, \bar{Z}_r)
\prod_{K=1}^{6h-6+2N}
   \left[ \int dz \wedge d\bar{z} \,i \left( \mu_{K}b
          +   
           \bar{\mu}_{K}\tilde{b}\right)
    \right].~~~~~~~~
\label{eq:Gammaghost}
\end{eqnarray}

In order to do so, we first rewrite the ghost path integral on the right 
hand side as follows. 
With the insertion of $\prod_{r=1}^{N} c\tilde{c}(Z_{r}, \bar{Z}_{r})$, 
the integration over $c$ and $\tilde{c}$
can be considered as the one over those $c, \tilde{c}$ which 
vanish at the punctures $z=Z_{r}$.
Let $\phi_\alpha$ $(\alpha =1,\ldots, 3-3h+N)$ be a basis
of the holomorphic quadratic differentials on 
the punctured Riemann surface $\Sigma$
which 
have no more than
simple poles at the punctures $Z_{r}$.
We decompose $b,\tilde{b}$
into the zero-modes and the nonzero-modes as
\begin{equation}
b(z, \bar{z})
=
\sum_\alpha \phi_\alpha (z)b_\alpha +b^\prime (z, \bar{z})\,,
\qquad
\tilde{b}(z, \bar{z})
=
\sum_\alpha \bar{\phi}_\alpha (\bar{z})\tilde{b}_\alpha 
+\tilde{b}^\prime (z, \bar{z})\,.
\end{equation}
The path integral measure $\left[db d\tilde{b} 
dcd\tilde{c}\right]_{g_{z\bar{z}}^\mathrm{A}}
\prod_{r=1}^{N} c\tilde{c}(Z_{r}, \bar{Z}_{r})$ can be rewritten as 
\begin{equation}
\left[db d\tilde{b} 
dcd\tilde{c}\right]_{g_{z\bar{z}}^\mathrm{A}}
\prod_{r=1}^{N} c\tilde{c}(Z_{r}, \bar{Z}_{r})
=
\left[db^\prime d\tilde{b}^\prime 
dcd\tilde{c}\right]_{g_{z\bar{z}}^\mathrm{A}}
\prod_\alpha \left(db_\alpha d\tilde{b}_\alpha\right)
\prod_{r=1}^{N} c\tilde{c}(Z_{r}, \bar{Z}_{r})
J[g_{z\bar{z}}^\mathrm{A}]
\,,
\end{equation}
with the Jacobian factor $J[g_{z\bar{z}}^\mathrm{A}]$. 
Then the path integral on the right hand 
side of eq.(\ref{eq:Gammaghost}) is equal to 
\begin{eqnarray}
&&
J[g_{z\bar{z}}^\mathrm{A}]
\int\prod_\alpha \left(db_\alpha d\tilde{b}_\alpha\right)
\prod_{K=1}^{6h-6+2N}
   \left[ \sum_\alpha \int dz \wedge d\bar{z} \,i 
   \left( \mu_{K}b_\alpha\phi_\alpha
          +   
           \bar{\mu}_{K}\tilde{b}_\alpha \bar{\phi}_\alpha
    \right)
    \right]
\nonumber
\\
& &
\qquad 
\times
\int\left[db^\prime d\tilde{b}^\prime 
dcd\tilde{c}\right]_{g_{z\bar{z}}^\mathrm{A}}
e^{-S^{bc}[b^\prime ,\tilde{b}^\prime ,c,\tilde{c}]}
\prod_{r=1}^{N}c\tilde{c}(Z_r, \bar{Z}_r)\,.
\end{eqnarray}
If we consider a ghost path integral of the form
\begin{equation}
\int \left[ db d\tilde{b} dc d\tilde{c} 
     \right]_{g^{\mathrm{A}}_{z\bar{z}}}
 e^{-S^{bc}} 
  \prod_{r=1}^{N}c\tilde{c} (Z_{r}, \bar{Z}_{r})
 \prod_{\alpha=1}^{3h-3+N} 
    b\tilde{b} (S_{\alpha}, \bar{S}_{\alpha})\,,
\label{eq:bc-correlator}
\end{equation}
for arbitrary $S_{\alpha}$ $(\alpha=1,\ldots,3h-3+N)$ instead, 
we can show that it is equal to  
\begin{eqnarray}
& &
J[g_{z\bar{z}}^\mathrm{A}]
\int\prod_\gamma \left(db_\gamma d\tilde{b}_\gamma\right)
\prod_{\beta=1}^{3h-3+N} 
\left[
    \sum_\alpha b_\alpha \phi_\alpha (S_{\beta})
    \sum_\alpha \tilde{b}_\alpha \bar{\phi}_\alpha(\bar{S}_{\beta})
\right]
\nonumber
\\
& &
\qquad
\times
\int\left[db^\prime d\tilde{b}^\prime 
dcd\tilde{c}\right]_{g_{z\bar{z}}^\mathrm{A}}
e^{-S^{bc}[b^\prime ,\tilde{b}^\prime ,c,\tilde{c}]}
\prod_{r=1}^{N}c\tilde{c}(Z_r, \bar{Z}_r)~.
\end{eqnarray}
Therefore we can see that the right hand side of eq.(\ref{eq:Gammaghost}) 
is rewritten as
\begin{eqnarray}
\lefteqn{
\frac{\det \left( \int dz \wedge d\bar{z} \,i\mu_{K} \phi_\alpha\ ,
                  \int dz \wedge d\bar{z} \,i\mu_{K} \bar{\phi}_\alpha
                  \right)}
         {\left| \det \phi_{\alpha} (S_{\beta}) \right|^{2}}
}
\nonumber
\\
& &
\times
\int \left[ db d\tilde{b} dc d\tilde{c} 
     \right]_{g^{\mathrm{A}}_{z\bar{z}}}
 e^{-S^{bc}} 
  \prod_{r=1}^{N}c\tilde{c} (Z_{r}, \bar{Z}_{r})
 \prod_{\alpha=1}^{3h-3+N} 
    b\tilde{b} (S_{\alpha}, \bar{S}_{\alpha})~.
\label{eq:muphiphiphi}
\end{eqnarray}

The bosonization 
formula~\cite{AlvarezGaume:1987vm,Verlinde:1986kw,
Dugan:1987qe,Sonoda:1987ra,D'Hoker:1989ae} 
implies that eq.(\ref{eq:bc-correlator}) is evaluated as 
\begin{eqnarray}
\lefteqn{
\int\left[dbd\tilde{b}dcd\tilde{c}\right]_{g_{z\bar{z}}^\mathrm{A}}
e^{-S^{bc}}
\prod_{r=1}^{N}c\tilde{c}(Z_r, \bar{Z}_r)
\prod_{\alpha =1}^{3h-3+N}b\tilde{b}(S_\alpha , \bar{S}_\alpha)
}
\nonumber
\\
&= &
Z^{X} [g_{z\bar{z}}^\mathrm{A}]
\left(\det \mathop{\mathrm{Im}} \Omega \right)^{\frac{1}{2}}
  \left| \theta [\xi^\prime ] (0|\Omega) \right|^{2}
  \prod_{r=1}^{N}\left(2g_{Z_r\bar{Z}_r}^\mathrm{A}\right)^{-1}
  \prod_{\alpha =1}^{3h-3+N}\left(2g_{S_\alpha\bar{S}_\alpha}\right)^2
\nonumber
\\
& &
\quad
\times
\exp
\left[
-\sum_{\alpha <\beta}G^\mathrm{A}(S_\alpha ; S_\beta )
-\sum_{r <s}G^\mathrm{A}(Z_r ; Z_s )
+\sum_{\alpha ,r}G^\mathrm{A}(S_\alpha ; Z_r )
\right]~,
\label{eq:bc-bosonization}
\end{eqnarray}
up to a numerical multiplicative constant,
where $\xi'$ is given by
\begin{equation}
\xi' \equiv \sum_{\alpha =1}^{3h-3+N} \int^{S_{\alpha}}_{P_{0}} \omega
        - \sum_{r=1}^{N} \int^{Z_{r}}_{P_{0}} \omega
        -3 \Delta
\qquad \pmod{\mathbb{Z}^{h} + \mathbb{Z}^{h} \Omega}~.
\label{eq:xi-prime}
\end{equation}
Now let us multiply the both sides of (\ref{eq:bc-bosonization}) by
\begin{eqnarray}
Z^{X}[g^{\mathrm{A}}_{z\bar{z}}]^{3}
&=& \left(\det \mathop{\mathrm{Im}} \Omega \right)^{-\frac{3}{2}}
  \left| \theta [\xi] (0|\Omega) \right|^{-2}
  \left| \det \omega_{j}(\hat{z}_{i}) \right|^{2}
      \prod_{i=1}^{h} 
              \left(2g^{\mathrm{A}}_{\hat{z}_{i}\bar{\hat{z}}_{i}} 
              \right)^{-1}
\nonumber \\
&& \ \times
      \exp \left[ \sum_{i<j} G^{\mathrm{A}} (\hat{z}_{i};\hat{z}_{j})
                  - \sum_{i} G^{\mathrm{A}} (\hat{z}_{i};\hat{w}) \right]\,,
\end{eqnarray}
derived from eq.(\ref{eq:ZX-delta}). 
Here $\hat{z}_{i}$ $(i=1,\cdots h)$
and $\hat{w}$ can be arbitrarily chosen. 
As was done in Ref.~\cite{Sonoda:1987ra}, we take
\begin{eqnarray}
\hat{z}_{i}
&=&
S_i \qquad \qquad (i=1,\cdots h)\,,
\nonumber
\\
\hat{w}
&=&
z_{2h-2+N}\,,
\nonumber
\\
S_{h+I'}
&=&
z_{I'} \qquad \qquad (I'=1,\ldots,2h-3+N)\,,
\end{eqnarray}
where $z_I~(I=1,\cdots ,2h-2+N)$ correspond to the interaction points of 
the light-cone diagram. 
Then $\xi'$ given in eq.(\ref{eq:xi-prime})
equals to
$\xi$ given in eq.(\ref{eq:zeta}) and we find
\begin{eqnarray}
\lefteqn{
 \int\left[ dbd\tilde{b} dc d\tilde{c}\right]_{g^{\mathrm{A}}_{z\bar{z}}}
e^{-S^{bc}} \prod_{r=1}^{N} c\tilde{c} (Z_{r}, \bar{Z}_{r}) 
\prod_{i=1}^{h} b\tilde{b} \left(S_{i}, \bar{S}_{i} \right)
\prod_{I'=1}^{2h-3+N}  b\tilde{b} (z_{I'}, \bar{z}_{I'} )
}
\nonumber \\
& = & Z^{X} [ g^{\mathrm{A}}_{z\bar{z}} ]^{-2}
    \frac{\left| \det \omega_{j} \left( S_{i} \right) 
          \right|^{2}}
         {\det \mathop{\mathrm{Im}} \Omega}
    \left| \partial^{2} \rho (z_{2h-2+N}) \right|^{-2}
    \prod_{i=1}^{h} \left| \partial \rho \left( S_{i} \right)
                    \right|^{2}
\nonumber \\
&& \quad 
    \times 
    e^{-2 (h-1)c}
    \prod_{r=1}^{N} \left( \left( 2g^{\mathrm{A}}_{Z_r\bar{Z}_r} \right)^{-1}
                           \left| \alpha_{r} \right|^{-1}
                    \right)
    \prod_{I=1}^{2h-2+N} \left( 2g^{\mathrm{A}}_{z_{I}\bar{z}_{I}}
                                \left| \partial^{2} \rho (z_{I}) \right|
                          \right)~,
\label{eq:bc-correlator2}
\end{eqnarray}
using eqs.(\ref{eq:metric-z-Arakelov}),
(\ref{eq:alpha-Arakelov}) and (\ref{eq:partial2rhozI}).

Following Ref.~\cite{D'Hoker:1987pr},
let us take a basis $\phi_{\alpha} = (\phi_{i},\phi_{h+I'})$
of the holomorphic quadratic differentials as
\begin{eqnarray}
\phi_{i} &=& d\rho \, \omega_{i} 
  \qquad (i=1,\ldots,h)~,
\nonumber \\
\phi_{h+I'} &=& d \rho \, \tilde{\omega}_{z_{0}-z_{I'}}
  \qquad (I'=1,\ldots,2h-3+N)~,
\label{eq:quadratic-complex}
\end{eqnarray}
where
$\tilde{\omega}_{P-Q}$ denotes the abelian differential
of the third kind with simple poles at $P$ and $Q$ of residues
$1$ and $-1$ and with purely imaginary periods, which is
given by
\begin{equation}
\tilde{\omega}_{P-Q}
 =  dz \partial_{z} \ln \frac{E(z,P)}{E(z,Q)}
      - 2\pi i \omega \frac{1}{\mathop{\mathrm{Im}}\Omega}
                   \mathop{\mathrm{Im}} \int^{P}_{Q} \omega~.
\end{equation}
For the basis~(\ref{eq:quadratic-complex})
we have
\begin{equation}
\phi_{i} (z_{I'}) =0~,
\qquad 
\phi_{h+I'} (z_{J'}) = \delta_{I',J'} \, \partial^{2} \rho (z_{I'})~,
\end{equation}
and thus
\begin{equation}
\det \phi_{\alpha} (S_{\beta})
  = \det \omega_{j} (S_{i})
    \prod_{I'=1}^{2h-3+N} \partial^{2} \rho (z_{I'})
    \prod_{i=1}^{h} \partial \rho (S_{i})~.
\label{eq:det-phi}
\end{equation}
In Ref.~\cite{D'Hoker:1987pr}, it is shown that
\begin{equation}
\det \left( \int dz \wedge d\bar{z} \,i\mu_{K} \phi_\alpha\ ,
                  \int dz \wedge d\bar{z} 
                    \,i \bar{\mu}_{K} \bar{\phi}_\alpha
                  \right)
\propto
\det \mathop{\mathrm{Im}} \Omega~,
\label{eq:muphiOmega}
\end{equation}
up to a numerical factor. 
Putting eqs.(\ref{eq:muphiphiphi}), (\ref{eq:bc-correlator2}), 
(\ref{eq:det-phi}), (\ref{eq:muphiOmega}) 
and (\ref{eq:expressionI-Gamma}) together, we obtain
eq.(\ref{eq:Gammaghost}).


\bibliographystyle{utphys}
\bibliography{SFTJune27_13.bib}

\end{document}